\documentclass[aps,reprint,superscriptaddress,longbibliography,prb,floatfix]{revtex4-2}
\usepackage{graphicx}
\usepackage{amsmath}

\usepackage{amsfonts}
\usepackage{comment}
\usepackage{bbm}
\usepackage{bm}
\usepackage{comment}
\usepackage{esint}
\usepackage{cancel}
\usepackage[dvipsnames]{xcolor}
\usepackage{microtype}
\usepackage[normalem]{ulem}
\usepackage{mathrsfs}
\usepackage{multirow}
\usepackage{array}
\usepackage{lipsum}
\usepackage[unicode=true, colorlinks=true, citecolor={blue!80!black}, urlcolor={blue!50!black}, linkcolor = {blue!80!black}]{hyperref}

\DeclareMathOperator{\sgn}{sgn}
\DeclareMathOperator{\re}{\mathrm{Re}}
\DeclareMathOperator{\im}{\mathrm{Im}}
\DeclareMathOperator{\tr}{\mathrm{Tr}}
\DeclareMathOperator{\pf}{\mathrm{Pf}}

\DeclareMathOperator{\diag}{diag}

\newcommand{\abs}[1]{\left\vert#1\right\vert}

\newcommand{\av}[1]{\left\langle#1\right\rangle}

\graphicspath{{../figures/}}

\begin{document}
\title{
Decoherence of Majorana zero modes mediated by gapless fermions
}
\author{Sauri Bhattacharyya}
\affiliation{Dipartimento di Fisica, Sapienza Università di Roma, Piazzale Aldo Moro 2, 00185 Rome, Italy}
\author{Marco Grilli}
\affiliation{Dipartimento di Fisica, Sapienza Università di Roma, Piazzale Aldo Moro 2, 00185 Rome, Italy}
\author{Bernard van Heck}
\affiliation{Dipartimento di Fisica, Sapienza Università di Roma, Piazzale Aldo Moro 2, 00185 Rome, Italy}

\date{\today}
\begin{abstract}
We study the decoherence of a collection of Majorana zero modes weakly coupled to a gapless reservoir of non-interacting fermions.
Using the Born-Markov approximation, we derive a Lindblad master equation for the dissipative dynamics of the Majorana zero modes.
Due to the long-range coupling between Majorana zero modes mediated by the gapless reservoir, the Lindblad jump operators are \emph{non-local} linear combinations of the Majorana operators.
We show that, as a consequence, the dissipative dynamics can exhibit long relaxation times, i.e. a slow decay of fermion parities.
A spectral analysis of the Liouvillian shows that the slow-down is suppressed as a power law of the distance between Majorana zero modes.
Finally, we validate the Lindblad equation by comparison with unbiased numerical simulations of the time evolution of the full density matrix.
In particular, these illustrate that non-Markovian dynamics establishes non-local correlations at small times.
\end{abstract}

\maketitle


\section{Introduction}

Majorana zero modes are bound states associated with low-dimensional superconductors in a topological phase~\cite{alicea2012,flensberg2021}.
Spatially, they are pinned to the ends of a superconducting wire~\cite{kitaev2001} or to the vortex cores in superconducting films~\cite{read2000,ivanov2001}.
Energetically, they are pinned to the Fermi level of the superconductor, at least provided that their separation is much larger than the superconducting coherence length.

Two well-separated Majorana zero modes form a non-local fermionic mode that can be emptied or occupied at no energy cost.
It follows that a collection of $N$ pairs of Majorana zero modes leads to a ground state degeneracy of $2^N$.
In the paradigm of topological quantum computation~\cite{nayak2008,dassarma2015}, the degenerate ground state space is used to encode qubits.
The non-locality of the encoding leads to the theoretical expectation that topological qubits based on Majorana zero modes will have long decoherence times~\cite{cheng2012}.

This expectation has the premise that the total fermion parity of the collection of Majorana zero modes is conserved. Indeed, the exchange of quasiparticles with the environment -- \emph{quasiparticle poisoning} -- would cause the decoherence of a quantum state encoded in the Majorana zero modes~\cite{goldstein2011,budich2012,rainis2012}.
This is why, although transport experiments probing the existence of Majorana zero modes require external metallic leads coupled to them, theoretical blueprints for Majorana-based quantum computing~\cite{alicea2011,hyart2013,aasen2016,vijay2016,karzig2017} exclude the presence of metallic leads altogether, or rely on the ability to turn off the coupling to them.

Not surprisingly, then, studies on the decoherence of quantum states encoded in Majorana zero modes~\cite{goldstein2011,budich2012,rainis2012,cheng2012,schmidt2012,konschelle2013,mazza2013,pedrocchi2015,hu2015,lai2018,knapp2018,munk2019,lai2020,karzig2021,alase2025} have focused on other sources of decoherence than the coupling to a metallic lead, such as charge noise or poisoning from out-of-equlilibrium quasiparticles populating the superconductor.
In contrast, with the exception of earlier work by Campbell~\cite{campbell2015}, the dissipative evolution of a collection of Majorana zero modes coupled to a gapless fermionic reservoir has not been studied in further detail.
In this article, we turn our attention to this very scenario and find instead that it has quite interesting features.

\begin{figure}[t!]
    \includegraphics{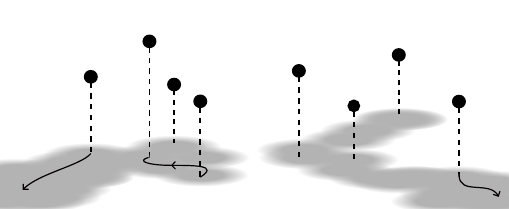}
    \caption{We consider a collection of Majorana zero modes (black dots) coupled by local tunneling (dashed lines) to a gapless reservoir of non-interacting fermions (gray). The reservoir supports trajectories linking different Majorana zero modes, as well as trajectories escaping away.}
    \label{fig:setup}
\end{figure}

Namely, we consider a collection of Majorana zero modes tunnel coupled to a generic reservoir of non-interacting fermions, see Fig.~\ref{fig:setup}. Adopting a standard open quantum systems approach~\cite{breuer2002}, we integrate out the microscopic degrees of freedom of the reservoir and derive a Lindblad master equation for the reduced density matrix describing the Majorana zero modes. When different Majorana zero modes are coupled via the reservoir, the jump operators entering the Lindblad equation are linear combinations of them and are, therefore, non-local. 

We analyze the consequence of these non-local couplings by a spectral analysis of the Liouville operator associated with the Lindblad equation~\cite{prosen2008,prosen2010}. We find that they can slow down the decay of the Majorana zero modes towards the fully mixed state of the degenerate ground state manifold, the natural steady state of the Lindblad equation.
In fine-tuned cases, the spectral gap of the Liouville operator can vanish, leading to the occurrence of additional steady states other than the fully mixed state.
More generically, the slow-down effects disappear as a power law of the length of the trajectories connecting pairs of Majorana zero modes via the reservoir.
Nevertheless, we argue that these effects are in principle observable in future experiments which can initialize and measure the parity encoded in different pairs of Majorana zero modes, e.g.~with the technique demonstrated in Ref.~\cite{microsoft2025}.

The paper is structured as follows.
In the next Section~\ref{sec:model}, after a brief description of the underlying microscopic Hamiltonian, we present the Lindblad model for a collection of Majorana zero modes coupled to a reservoir, and discuss its main features.
In Sec.~\ref{sec:decay} we analyze the Liouvillian time evolution of Gaussian states, solving it explicitly in some simple limits that reveal the slowdown decay of fermion parity.
In Sec.~\ref{sec:spectrum} we carry out a spectral analysis of the Liouvillian of the model using Prosen's third quantization method. We treat explicitly the case of a free-electron gas reservoir, studying the dependence of the results on the dimensionality of the reservoir and on the distance between Majorana zero modes.
In Sec.~\ref{sec:numerics}, we compare the time evolution obtained from the Lindblad equation with unbiased numerical simulations of the von Neumann equation of the system plus the reservoir, in this case a critical Kitaev chain.
The comparison illustrates both the validity of the Lindblad approach and its limitations: in particular, the delicateness of the Markov approximation to describe the coupling between distant Majorana zero modes.
We conclude in Sec.~\ref{sec:conclusions} discussing possible consequences, extensions and improvements of our findings.
Long or technical derivations are relegated to Appendices.

\section{Model}
\label{sec:model}

\subsection{Microscopic Hamiltonian for Majorana zero modes coupled to a fermionic reservoir}

We consider $2N$ Majorana zero modes with Hermitian operators $a_i$ obeying
\begin{equation}
\{a_i, a_j\}=\delta_{ij}\,,
\end{equation}
for $i, j=1,\dots, 2N$. We assume that the existence of Majorana zero modes is supported by one or more topological superconductors of which they constitute the relevant low-energy degrees of freedom. Their positions in real space are determined by the geometry of the underlying superconductors as well as the location of vortices or defects. We assume that the Majorana zero modes are all well separated, so that there are no direct couplings between them due to the overlap of their wave functions through the topological superconductors.

We consider a situation in which the Majorana zero modes (or at least a subset of them) are coupled to a fermionic reservoir.
We assume that the fermionic reservoir is well described by a non-interacting Hamiltonian, quadratic in the fermionic operators, possibly including pairing terms.
To treat pairing terms on the same footing as number-conserving ones, it is convenient to adopt a Majorana basis to describe the reservoir.
For this purpose we introduce a second set of Majorana operators $b_n$.
We further assume that the Majorana zero modes are locally coupled to the reservoir by tunneling terms.
The Hamiltonian of the model is therefore the sum of two quadratic terms:
\begin{align}\label{eq:microscopic_model}
H = \frac{i}{2}\sum_{in} W_{in} a_i b_n + \frac{i}{2} \sum_{nm} B_{nm}b_nb_m\,.
\end{align}
The first term describes tunneling between the Majorana zero modes and the reservoir, with tunnel couplings $W_{in}$.
The second term is the Hamiltonian of the reservoir written in the Majorana basis, with an antisymmetric matrix $B_{nm}=-B_{mn}$. The reservoir Hamiltonian can possibly describe multiple disconnected reservoirs, a situation that we will at times consider.

In the remainder of the paper we will consider explicitly some specific models for the reservoir and the tunneling Hamiltonian, giving expressions for the parameters that appear in Eq.~\eqref{eq:microscopic_model}.
For now we only wish to emphasize the basic assumptions of the model:
first, that the Majorana zero modes are only coupled to the reservoirs and not directly to each other, so that the Hamiltonian does not contain quadratic terms like $a_i a_j$. This assumption implies that the coherence length in the topological superconductor(s) hosting the Majorana zero modes is much shorter than the distance between any two of them. Second, we also neglect the presence of charging energy terms in any of the superconductors, so that there are no interacting terms proportional e.g. to $a_ia_ja_ka_l$. Similarly, any interactions within the reservoir and between the reservoir and the Majorana zero modes are neglected.

Despite its simplicty, the quadratic model of Eq.~\eqref{eq:microscopic_model} would be a reasonable starting point to study several heterostructures of experimental interest.
For instance, it could describe a network of topological superconducting wires which includes normal sections and leads, such as those engineered starting from two-dimensional electron gases~\cite{flensberg2021}.
It could also describe a set of vortices in a 2D topological superconductor coupled to the chiral edge modes present in a finite sample.
Finally, it could also describe Majorana zero modes engineered on the proximitized surface of a topological insulator~\cite{fu2008}, coupled to a residual mid-gap density of states in the bulk of the topological insulator. These examples also show that it is worthwhile to consider fermionic reservoirs of different dimensionality.

\subsection{Lindblad equation for the Majorana zero modes}

The state of the Majorana zero modes is described by a density matrix $\rho$ which acts in the Hilbert space spanned by the Majorana operators $a_i$, which has dimension $2^N$. Under the usual assumptions that the coupling to the reservoir is weak, that the reservoir is initially at thermal equilibrium at temperature $T$, and that dynamics of the reservoir is Markovian, we derive the following master equation for the density matrix:
\begin{subequations}\label{eq:lindblad}
\begin{equation}
\dot \rho = \mathcal{L}(\rho)
\end{equation}
with the Liouvillian $\mathcal{L}$ given by
\begin{align}\label{eq:liouvillian}
\mathcal{L}(\rho) &= -\frac{1}{4}\sum_{ij} \Lambda_{ij} [a_ia_j,\rho] + \mathcal{D}(\rho)\,,\\
\mathcal{D}(\rho) &=- \frac{\pi}{4} \sum_i J_{ii} \rho + \frac{\pi}{2}\sum_{ij} J_{ij}a_i \rho a_j\,.
\end{align}
\end{subequations}
The first term in Eq.~\eqref{eq:liouvillian} describes the unitary dynamics induced by the reservoir (the Lamb shift). The dissipator $\mathcal{D}(\rho)$ describes the dissipative part of the evolution induced by the reservoir.
The Liouvillian of Eq.~\eqref{eq:liouvillian} falls into the class of free fermionic Liouvillians.
Namely, the Hamiltonian generating the unitary dynamics is quadratic in the Majorana operators, while the Lindblad jump operators are linear.

Explicit expressions for the couplings entering the Lindblad equation can be given as spectral sums over the reservoir:
\begin{subequations}\label{eq:Lambda_ij_J_ij}
\begin{align}\label{eq:Lambda_ij}
\Lambda_{ij} &= \frac{1}{2}\sum_k \zeta_k^{-1} A_{ijk}\,,\\
J_{ij} &= \frac{1}{2}\sum_k \delta(\zeta_k) S_{ijk}\,,
\end{align}
\end{subequations}
where the index $k$ runs over all single-particle excitations of the reservoir, $\zeta_k$ are the corresponding excitation energies, and finally $S_{ijk}$ and $A_{ijk}$ are symmetric and anti-symmetric combinations of tunneling amplitudes between the Majorana zero modes and the reservoir.
Their detailed expressions are tedious to write and so are postponed to App.~\ref{app:lindblad}, see Eqs.~\eqref{eqs:S_and_A}.
The equations above are valid only for reservoirs with a continuous energy spectrum, so the sums have to be implemented as integrals, with the principal value prescription for Eq.~\eqref{eq:Lambda_ij}.

The couplings $\Lambda_{ij}$ and $J_{ij}$ are \emph{real-valued} matrix elements of $2N \times 2N$ matrices $\Lambda$ and $J$ with the properties
\begin{subequations}
\begin{align}
\Lambda&=-\Lambda^T\,,\\
J&=J^T\,.
\end{align}
\end{subequations}
Moreover, $J$ is a positive-definite matrix: all its eigenvalues are real and positive.

The derivation of the Lindblad equation from the microscopic Hamiltonian~\eqref{eq:microscopic_model} is carried out in App.~\ref{app:lindblad}.
It adheres quite closely to textbook derivations of master equations from microscopic models~\cite{breuer2002}, with the specificity that it is carried out entirely in terms of Majorana operators.
A similar weak-coupling derivation for a free fermionic model was also provided in Ref.~\cite{campbell2015}.
The properties of $\Lambda$ and $J$ are derived by construction in App.~\ref{app:lindblad}.

\subsection{Remarks on the model}

Before studying its solution, it is useful to make some general remarks on Eq.~\eqref{eq:lindblad}.

\subsubsection{Degeneracy and temperature}

An important assumption in the derivation of Eq.~\eqref{eq:lindblad} is that the Majorana zero modes are only coupled to each other via the reservoir. 
Thus, if the coupling to the reservoir is continuously sent to zero, the Majorana zero modes are all perfectly degenerate at zero energy.
In turn, this implies that the unitary part of the dynamics is entirely generated by the Lamb shift $\Lambda$.

There are three other consequences of this assumption.
The first, and more technical one, is that no secular approximation must be invoked to cast the Born-Markov approximation in Lindblad form, as shown in App.~\ref{app:lindblad}.
The second is that the matrix $J$ is purely real, even though the most generic Lindblad jump operators linear in Majorana operators could feature complex couplings.

The third consequence is that the couplings $J_{ij}$ do not depend on the temperature of the reservoir.
Hence, as pointed out by Campbell in Ref.~\cite{campbell2015}, the Markovian dynamics of the Majorana zero modes is independent of the temperature of the reservoir.
We interpret this property as a consequence of the particle-hole symmetry of Majorana zero modes, which couple symmetrically to electrons and holes and so are insensitive to the Fermi-Dirac distribution of the occupation numbers of the modes in the reservoirs.

In fact, the condition of perfect degeneracy can be relaxed to a more gentle condition that the temperature of the reservoir be larger than any direct couplings between Majorana zero modes.
This condition extends to the Lamb shift term themselves: we will see in the numerical simulations of Sec.~\ref{sec:numerics} that if $k_BT \lesssim \Lambda_{ij}$ the exact time evolution and the steady state  become temperature-dependent, unlike those predicted by the Lindblad equation~\eqref{eq:lindblad}.

\subsubsection{Gapped vs gapless reservoir}

The derivation of Eq.~\eqref{eq:lindblad} does not make any assumption on the presence of an energy gap at the Fermi level in the reservoir.
It is immediate to see that the dissipative couplings $J_{ij}$ vanish if there is a gap.
In this case, the Markovian dynamics induced by the reservoir is purely unitary and generated by the Lamb shift $\Lambda$.
Moreover, the couplings $\Lambda_{ij}$ will generically be exponentially suppressed in the distance between the Majorana zero modes.

In this respect,  Eq.~\eqref{eq:Lambda_ij_J_ij} yields results consistent with expressions existing in the literature, e.g. for the coherent coupling between Majorana zero mode via a gapped superconductor~\cite{zyuzin2013}.
Decoherence effects due to a bath of gapped fermions has been analyzed in detail in Ref.~\cite{goldstein2011}.
In the rest of this work, we will be focused on gapless reservoirs. 

\subsubsection{Non-locality}

Loosely speaking, $\Lambda_{ij}$ and $J_{ij}$ are the real and imaginary parts of the zero-energy Green's function propagating a fermion from the location of $a_i$ to the location of $a_j$ via the reservoir.
Thus, in general, both couplings will be non-zero if the reservoir is gapless and supports trajectories linking the locations of $a_i$ and $a_j$ (see Fig.~\ref{fig:setup}).
In the case of disconnected reservoirs coupling only to a single Majorana zero mode, characteristic of many transport experiments, there are no such linking trajectories and therefore only the diagonal couplings $J_{ii}$ are non-zero in Eq.~\eqref{eq:lindblad}.
In this case, as we show explicitly in the next Section, the time evolution predicted by the Lindblad equation is an exponential decay of all fermion parity observables towards zero, dictated only by local tunneling rates.

We are more interested in the case of a connected reservoir with linking trajectories.
Due to the absence of a gap, the reservoir will mediate long-range couplings between the Majorana zero modes, and the matrices $\Lambda_{ij}$ and $J_{ij}$ will be dense.
We will see in some simple models that, indeed, their off-diagonal matrix elements will generically decrease as a power law of the distance between Majorana zero modes, so that in limit of very large distances the case of disconnected reservoirs is recovered.

In this scenario, the dense positive matrix $J$ is diagonalized by a non-trivial orthogonal matrix $O$, $O J O^T = \diag(\kappa_1, \dots, \kappa_{2M})$ with eigenvalues $\kappa_i$ such that
\begin{equation}
\kappa_1 \geq \kappa_2 \geq \dots \geq \kappa_{2M} \geq 0\,.
\end{equation}
If we introduce a new set of Majorana operators
\begin{equation}
  \gamma_i=\sum_j O_{ij} a_j \,,
\end{equation}
the dissipator takes the form:
\begin{equation}
\mathcal{D}(\rho) = -\frac{\pi}{4}\sum_{i}\kappa_i\,(\rho + 2\gamma_i\rho\gamma_i)\,.
\end{equation}
In this basis the dissipative part appears as a sum over independent terms, each corresponding to a jump operator $\sqrt{2\pi\kappa_i} \gamma_i$. Although in practice we will not make use of this basis, it makes it clear that, unless the off-diagonal matrix elements of $J$ all vanish: (i) the jump operators are non-local linear combinations of the original Majorana operators; (ii) the decay towards the steady state will not be dictated by local tunneling rates only.

\subsubsection{Validity of the weak coupling approximation}

The Lindblad master equation is based on the Born-Markov approximation, thus on a perturbative expansion valid for weak couplings between the Majorana zero modes and the bath.
For the approximation to be valid, to begin with the couplings $W_{in}$ in Eq.~\eqref{eq:microscopic_model} must be much smaller than the topological superconducting gap in the system hosting the Majorana zero modes. Otherwise, the Hamiltonian~\eqref{eq:microscopic_model} would not be a good starting point to model the system-reservoir interaction.
Furthermore, as emphasized in recent literature on open quantum systems~\cite{nathan2020,mozgunov2020}, the validity of the Lindblad equation rests on the comparison between the bath correlation time on one hand and the coupling strength between the system and the reservoir on the other hand.

In particular, it is clear that the long-range couplings between Majorana zero modes appearing in the Lindblad equation cannot be established instantaneously.
Thus, we must require that the inverse tunneling rate be much longer than the time it takes to establish a correlation between distant Majorana zero modes via the reservoir.
There are, indeed, multiple two-point bath correlation functions that must be evaluated in order to derive the Lindblad equation - one for each non-local term in the dissipator, as detailed in App.~\ref{app:lindblad}.

Because the correlation time will increase with the distance between Majorana zero modes, broadly speaking the larger the distance, the weaker the tunnel couplings need to be in order for the Lindblad equation to be valid.
Increasing the coupling strengths between the Majorana zero modes and the reservoir or the distance between Majorana zero modes, non-Markovian effects become more relevant, causing larger deviations from the predictions of the Lindblad equation.
These aspects are further addressed in the exact numerical calculations of Sec.~\ref{sec:numerics}, see in particular Fig.~\ref{fig:plateau_scaling}, and in the derivation of App.~\ref{app:lindblad}.

\subsubsection{Steady states and spectral gap}

It is useful at this point to recall some basic notions about the spectral and steady-state properties of the Lindblad equation, following Ref.~\cite{costa2023} and referring to the literature, e.g. Ref.~\cite{albert2014}, for a general analysis.
We will assume, and in practice we will always find this assumption to be correct, that the Liouvillian of Eq.~\eqref{eq:liouvillian} can be diagonalized.
That is, one can find a complete set of eigenmodes $\rho_\alpha$ such that $\mathcal{L}(\rho_\alpha) = -\lambda_\alpha \rho_\alpha$.
The positivity of $J$ guarantees that the eigenvalues $\lambda_\alpha$ all have positive real part: $\re(\lambda_\alpha)\geq 0$.

Then, for an arbitrary initial density matrix given as a linear combination of the eigenmodes,
\begin{equation}\label{eq:general_initial_condition}
\rho(0) = \sum_\alpha c_\alpha \rho_\alpha\,,
\end{equation}
the master equation~\eqref{eq:lindblad} leads to a completely positive and trace preserving quantum map at all times
\begin{equation}
\rho(t) = \sum_\alpha c_\alpha e^{-\lambda_\alpha t} \rho_\alpha\,.
\end{equation}

Because the trace of $\rho(t)$ has to be equal to one at all times, at least one steady-state density matrix with eigenvalue $\lambda_0=0$ can be found. In the case of Eq.~\eqref{eq:lindblad}, this is the fully mixed state
\begin{equation}\label{eq:fully_mixed_state}
\rho_0 = 2^{-M}\, \mathbb{I}\,.
\end{equation}
with $\mathbb{I}$ the identity operator. The property $\mathcal{L}(\rho_0)~=~0$ can be verified by direct substitution of $\rho_0$ into Eq.~\eqref{eq:liouvillian}.

The rest of the spectrum of $\mathcal{L}$ determines the rapidity with which the steady state is reached for a given initial state. Of particular importance is the \emph{spectral gap} of the Liouvillian,
\begin{equation}\label{eq:gap}
\Delta = \min_{\alpha \neq 0} \re(\lambda_\alpha)\,.
\end{equation}
which dictates the minimum possible rapidity, i.e. the relaxation rate at long times.
We will see that off-diagonal matrix elements $J_{ij}$ present in Eq.~\eqref{eq:liouvillian} tend to decrease the spectral gap.
In some fine-tuned cases, even a vanishing gap is possible, signifying the occurrence of additional steady states beyond the fully mixed state.

\subsection{Time evolution of Gaussian states}
\label{subsec:gaussian_states}

While the Lindblad equation is valid for an arbitrary initial state of the Majorana zero modes, we will from now on assume that the initial state is Gaussian. Such a state
$\rho$ is completely specified by the covariance matrix~\cite{terhal2002,bravyi2004,bravyi2011,bravyi2017}:
\begin{equation}
\label{eq:covariance matrix}
M_{ij} = -i \tr\left(\rho [a_i, a_j]\right)\,.
\end{equation}
The covariance matrix is a real antisymmetric matrix. The off-diagonal matrix are the expectation values of the fermion parity operators $-2i a_i a_j$, taking values in the interval $[-1,1]$.

Given the covariance matrix, arbitrary expectation values of Majorana operators can be computed using Wick's theorem:
\begin{equation}\label{eq:wick}
\tr(a_i a_j\dots a_n) = \pf[\tfrac{i}{2}M_{(ij\dots n)}]\,.
\end{equation}
Here, the right hand side denotes the Pfaffian of the minor of the covariance matrix selected by the indices $i, j, \dots, n$ entering the expectation value on the left hand side.

As a real antisymmetric matrix, $M$ can be brought into a canonical block-diagonal form by an orthogonal transformation $R$,
\begin{equation}
R M R^T = \bigoplus_{p=1}^M \begin{bmatrix}
    0 & \lambda_p\\ -\lambda_p & 0
\end{bmatrix}\,.\\
\end{equation}
The $\lambda_p$ are real numbers in the interval $[-1, 1]$. The Gaussian state corresponding to such a covariance matrix can be written as
\begin{equation}
\rho = \frac{1}{2^M}\prod_{p=1}^M \left(\mathbb{I}-2i\lambda_p \tilde{a}_{2p-1}\tilde{a}_{2p}\right)\,,
\end{equation}
with $\tilde{a} = R a$. We note that the Gaussian state only contains terms with an even number of Majorana operators. This property is desirable because terms in the density matrix involving an odd number of Majorana operators correspond to coherences between the even and odd fermion parity sectors of the Hilbert space. Their presence should be discarded to enforce the fermion parity super-selection rule.

Since the Liouvillian is free, Gaussian states remain Gaussian under time evolution.
Thus, all the information about the system can be contained in a time-dependent covariance matrix $M(t)$.
The time evolution of the covariance matrix follows from the Lindblad equation and Wick's theorem, and takes the compact form
\begin{equation}\label{eq:dotM}
\dot M = \frac{1}{2}\,[\Lambda, M]- \frac{\pi}{2}\,\{J, M\}\,.
\end{equation}
A derivation of this result is contained in App.~\ref{app:covariance_matrix_evolution}.
Solving Eq.~\eqref{eq:dotM} is technically more simple and physically more transparent than solving Eq.~\eqref{eq:lindblad}.
In the next Section, we solve Eq.~\eqref{eq:dotM} in some simple cases in order to illustrate the physical consequences of the non-local couplings in the Lindblad equation.\\

\section{Occurrence of slow parity decay}
\label{sec:decay}

Let us consider the case in which each Majorana zero mode is coupled to a different reservoir, so that there are no trajectories linking different modes (see Fig.~\ref{fig:setup}).
In this case, $\Lambda=0$ and $J$ is a purely diagonal matrix.
Eq.~\eqref{eq:dotM} reduces to a set of decoupled equations, one for each element of the covariance matrix:
\begin{equation}
\dot{M}_{ij} = -\frac{\pi (J_{ii}+J_{jj})}{2}\,M_{ij}\,.
\end{equation}
The result is a simple exponential decay towards the fully mixed state with $M=0$,
\begin{equation}
M_{ij}(t) = e^{-\pi(J_{ii}+J_{jj})t/2}\, \,M_{ij}(0)\,.
\end{equation}
This time evolution is simple to interpret: the decay time for the fermion parity $-2ia_i a_j$ is the sum of the tunneling rates coupling the reservoirs to the Majorana zero modes $i$ and $j$.
We recover the fact that the fermion parities encoded in the Majorana zero modes decay exponentially with a relaxation rate set by the local tunneling rates between each Majorana and the reservoir.

\begin{figure}
    \includegraphics{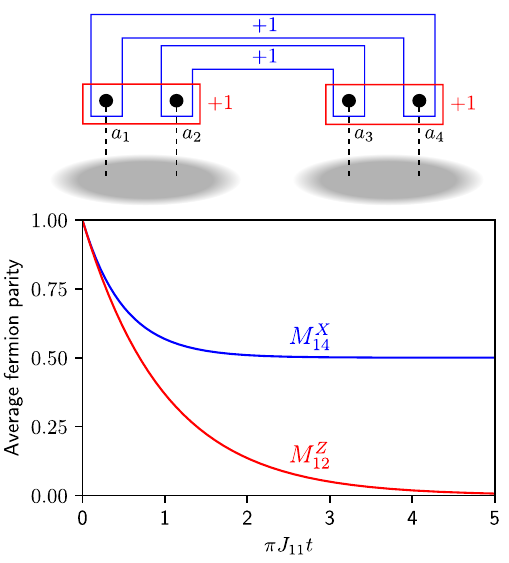}
    \caption{Four Majorana zero modes coupled pairwise to two different reservoirs. The red and blue boxes represent two alternative ways to initialize fermion parity, denoted Z-type and X-type in the main text. The decay of fermion parity can be much slower for X-type initial conditions (blue) than for Z-type (red).}
    \label{fig:pairwise}
\end{figure}

We now show that this conventional wisdom does not adequately capture the case of a connected reservoir linking different Majorana zero modes.
To do so, take the minimal example of Fig.~\ref{fig:pairwise}: we consider four Majorana zero modes pairwise coupled to two different reservoirs.
With this constraint, the $\Lambda$ and $J$ matrices assume a block-diagonal form, without any term coupling $a_1$ and $a_2$ to $a_3$ and $a_4$. In this case, Eq.~\eqref{eq:dotM} yields the following system of equations:
\begin{widetext}
\begin{align}\label{eq:dotM_example}
\begin{bmatrix}
    \dot{M}_{12} \\
    \dot{M}_{13} \\
    \dot{M}_{14} \\
    \dot{M}_{23} \\
    \dot{M}_{24} \\
    \dot{M}_{34}
\end{bmatrix}=-\frac{1}{2}
\begin{bmatrix}
\pi(J_{11}+J_{22}) & 0 & 0 & 0 & 0 & 0 \\
0 & \pi(J_{11}+J_{33}) & \pi J_{34}-\Lambda_{34} & \pi J_{12}-\Lambda_{12} & 0 & 0 \\
0 & \pi J_{34}+\Lambda_{34} & \pi(J_{11}+J_{44}) & 0 & \pi J_{12}-\Lambda_{12} & 0 \\
0 & \pi J_{34}+\Lambda_{12} & 0 & \pi(J_{22}+J_{33}) & \pi J_{34}-\Lambda_{34} & 0 \\
0 & 0 & \pi J_{12}+\Lambda_{12} & \pi J_{34}+\Lambda_{34} & \pi(J_{22}+J_{44}) & 0 \\
0 & 0 & 0 & 0 & 0 & \pi(J_{33}+J_{44}) \\
\end{bmatrix}
\begin{bmatrix}
    M_{12} \\
    M_{13} \\
    M_{14} \\
    M_{23} \\
    M_{24} \\
    M_{34}
\end{bmatrix}
\end{align}
\end{widetext}
We see that $M_{12}$ and $M_{34}$ evolve as in the case of disconnected reservoirs, while the time evolution of the other four parity operators are intertwined by the off-diagonal matrix elements of $\Lambda$ and $J$.

Let us contrast the resulting time evolution for the two different initial conditions illustrated in Fig.~\ref{fig:pairwise}.
Explicitly, these are
\begin{subequations}
\label{eq:MX_MZ}
\begin{equation}
M^Z =
\begin{bmatrix}
0 & +1 & 0 & 0 \\
-1 & 0 & 0 & 0 \\
0 & 0 & 0 & +1 \\
0 & 0 & -1 & 0
\end{bmatrix}
\end{equation}
and
\begin{equation}
M^X =
\begin{bmatrix}
0 & 0 & 0 & +1 \\
0 & 0 & +1 & 0 \\
0 & -1 & 0 & 0 \\
-1 & 0 & 0 & 0
\end{bmatrix}\,.
\end{equation}
\end{subequations}
We will also refer as these two possible initial conditions respectively as $Z$-type (red boxes in Fig.~\ref{fig:pairwise}) and $X$-type (blue boxes in Fig.~\ref{fig:pairwise}).
Both initial conditions correspond to pure states of even total parity.
However, in the Z-type initial condition, the state of the system is initialized in the joint parity eigenstate of $-2ia_1a_2$ and $-2ia_3a_4$ with eigenvalues $+1$.
On the other hand, in the X-type initial condition, the state of the system is initialized in the joint parity eigenstate of $-2ia_1a_4$ and $-2ia_2a_3$ with eigenvalues +1.
The important distinction between them is that the $Z$-type initial condition does not mix Majorana zero modes coupled to different reservoirs, while the $X$-type does.

For simplicity, we discuss the results for a maximally symmetric choice of parameters: 
\begin{equation}\label{eq:symmetric_parameter_choice}
J_{11}=J_{22}=J_{33}=J_{44}\,,\quad J_{12}=J_{34}\,,\quad \Lambda_{12}=\Lambda_{34}\,.
\end{equation}
The eigenvalues of Eq.~\eqref{eq:dotM_example} for this choice of parameters are $\pi J_{11}$ (four times) and
\begin{equation}
\pi J_{11} \pm \sqrt{\pi^2 J_{12}^2-\Lambda_{12}^2} \equiv \pi J_{11}\pm \Delta_{12}\,.
\end{equation}
We see that the behavior of the eigenvalues is controlled by the single quantity $\Delta^2_{12}=\pi^2 J_{12}^2-\Lambda_{12}^2$. The case $\Delta^2_{12}=0$ is an exceptional point which separates two distinct behaviors.

On one hand, if $\Delta^2_{12}>0$, the eigenvalues are purely real, but they are repelled from the point $\pi J_{11}$ which corresponds to the case of disconnected reservoirs. In particular, the spectral gap is given by $\pi J_{11}-\Delta_{12}$ and can approach zero in the limiting case $\Lambda_{12}=0$ and $J_{12}\to J_{11}$ (note that for this symmetric choice of parameters there is a bound $J_{12}\leq J_{11}$ due to the positivity of the matrix $J$). On the other hand, if $\Delta^2_{12}<0$, the eigenvalues acquire an imaginary part due to the Lamb shift $\Lambda_{12}$ while the spectral gap remains at $\pi J_{11}$.

The observable consequences of these two scenarios become visible by comparing the time evolution of $Z$-type and $X$-type initial conditions (see also Fig.~\ref{fig:pairwise}). For the $Z$-type initial conditions, the time evolution is as simple as
\begin{equation}\label{eq:Ztype_evolution}
M^Z(t) =  e^{-\pi J_{11} t}\,M^Z(0)\,.\\
\end{equation}
It mirrors the case of disconnected reservoirs: we do not see the effect of off-diagonal terms in the dissipator or the presence of the Lamb shift $\Lambda_{12}$.

On the other hand, $X$-type initial conditions have the following time dependence:
\begin{subequations}
\label{eq:Xtype_evolution}
\begin{align}
M^X_{12}(t) &= 0\\
M^X_{13}(t) &= -e^{-\pi J_{11}t} \,r\,\sinh(\Delta_{12}\,t) \\
M^X_{14}(t) &= e^{-\pi J_{11}t}\,\cosh(\Delta_{12}\,t)\\
M^X_{23}(t) &= e^{-\pi J_{11}t}\,\cosh(\Delta_{12}\,t) \\
M^X_{24}(t) &= -e^{-\pi J_{11}t} \,r^{-1}\,\sinh(\Delta_{12}\,t) \\
M^X_{34}(t) &=0\,.
\end{align}
\end{subequations}
with $r^2=(\pi J_{12}+\Lambda_{12})/(\pi J_{12}-\Lambda_{12})$.

We see that in this case the time dependence is more interesting. If $\Delta_{12}$ is purely imaginary, the system decays towards the fully mixed state, but with an overdamped oscillation rather than a pure exponential decay. On the other hand, if $\Delta^2_{12}>0$, the decay of fermion parities $M_{14}$ and $M_{23}$ has a fast component, with a decay rate $\pi J_{11}+\Delta_{12}$, and a slow component, with a decay rate $\pi J_{11}-\Delta_{12}$. In the limiting case in which the spectral gap vanishes, the slow component does not decay at all (blue curve in Fig.~\ref{fig:pairwise}). This signals the presence of a second steady state beyond the fully mixed state $\rho_0$. Indeed it can be checked that the covariance matrix
\begin{equation}
M = \frac{1}{2}\begin{bmatrix}
0 & 0 & -1 & +1 \\
0 & 0 & +1 & -1 \\
+1 & -1 & 0 & 0 \\
-1 & +1 & 0 & 0
\end{bmatrix}
\end{equation}
is a zero mode of Eq.~\eqref{eq:dotM} [or equivalently Eq.~\eqref{eq:dotM_example}] for this particularly symmetric parameter choice. This covariance matrix corresponds to the steady state
\begin{equation}
\rho = \frac{1}{4}\left(\mathbb{I}+ia_1a_3-ia_1a_4-ia_2a_3+ia_2a_4\right)\,,
\end{equation}
a balanced mixture of two pure states of opposite fermion parity.
For completeness, we note that the time evolution of the \emph{total} fermion parity is the same in both X-type and Z-type initial conditions,
\begin{equation}
\tr(-4 a_1 a_2 a_3 a_4) = e^{-2\pi J_{11}t}\,,
\end{equation}
so that a measurement of the total parity alone would not reveal the presence of long relaxation times or of a second steady state.

This simple example shows that the off-diagonal matrix elements in the dissipator can slow down the decay of fermion parity, at least for certain initial states.
Physically, this is a coherent effect related to the propagation of fermionic quasiparticles between two Majorana zero modes.
To determine whether this effect occurs generically or only in fine-tuned cases, and also to study the magnitude of the effect, in the next Section we turn to a more systematic study of the spectral gap of the Liouvillian.

\section{Spectral analysis of the Liouvillian}
\label{sec:spectrum}

The spectral gap of a free fermionic Liouvillian can be efficiently studied using the formalism of third quantization introduced by Prosen~\cite{prosen2008,prosen2010}.
Indeed, the Liouvillian can be represented as the following quadratic form~\footnote{We note that, in the most general case, the third-quantized Liouvillian features anomalous pairing terms proportional to $\mathcal{C}_i^\dagger \mathcal{C}_j^\dagger$, which are absent in Eq.~\eqref{eq:liouvillian_tq}. This simplification is a consequence of $J$ being a real matrix.}:
\begin{equation}\label{eq:liouvillian_tq}
\mathcal{L} = -\frac{1}{2}\sum_{ij}(\Lambda_{ij}+\pi J_{ij})\,\mathcal{C}^\dagger_i \mathcal{C}_j\,,
\end{equation}
where $\mathcal{C}_i$ and $\mathcal{C}^\dagger_i$ are creation and destruction super-operators. Essentially, $\mathcal{C}_i$ removes the Majorana operator $a_i$ from a string if it is present, and returns zero if it is absent. On the other hand, $\mathcal{C}^\dagger_i$ adds $a_i$ to a string if it is absent and returns zero if it is present.

\begin{figure*}[t!]
    \includegraphics{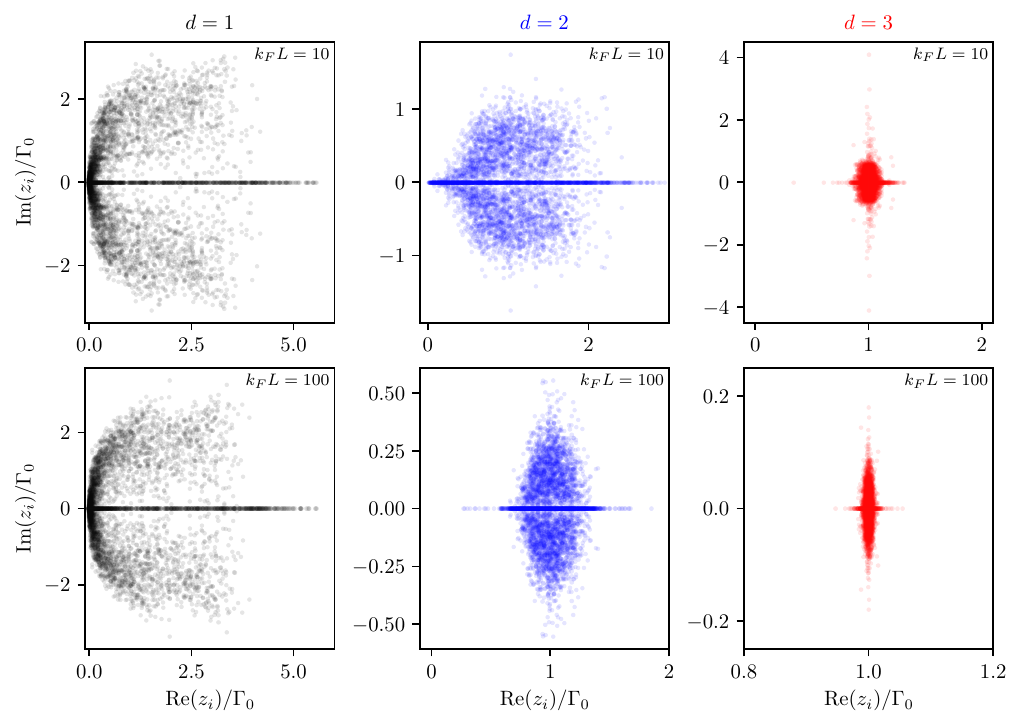}
    \caption{Spectra of the Liouvillian for a free-electron gas reservoir in $d=1,2,3$ dimensions, for ten Majoranas placed randomly in a hypercube of size $L$, smaller in the top row ($k_FL=10)$ and larger in the bottom row ($k_FL=100$). Note that in $d=2$ and $d=3$ panel, the range of the axes shrinks in the bottom row. }
    \label{fig:spectra}
\end{figure*}

Importantly, these operators obey canonical anti-commutation rules:
\begin{subequations}
\begin{align}
\{\mathcal{C}_i,\mathcal{C}^\dagger_j\}&=\delta_{ij}\,\\
\{\mathcal{C}_i,\mathcal{C}_j\}&=\{\mathcal{C}^\dagger_i,\mathcal{C}^\dagger_j\}=0\,.
\end{align}
\end{subequations}

The canonical structure made apparent in third quantization implies that the spectrum of the Liouvillian can be found by diagonalizing the matrix
\begin{equation}
Z = \Lambda + \pi J\,.
\end{equation}
The eigenvalues $z_i$ of $Z$ are either purely real or come in complex-conjugate pairs.
In either case, by the positivity of $J$, they all have positive real parts, $\re(z_i)\geq 0$, guaranteeing exponential decay rather than amplification of the initial state.
The spectrum of $\mathcal{L}$ is obtained by the sums of all possible subsets of the spectrum of $Z$.
Thus, in particular, the spectral gap defined in Eq.~\eqref{eq:gap} can be computed as
\begin{equation}
\Delta = \min_i \re(z_i)\,.
\end{equation}
As an example, the case of pairwise connected reservoirs explicitly treated in Eq.~\eqref{eq:dotM_example} of the previous section corresponds to the matrix
\begin{equation}
Z = \begin{bmatrix}
\pi J_{11} & \pi J_{12}+\Lambda_{12} & 0 & 0 \\
\pi J_{12}-\Lambda_{12} & \pi J_{22} & 0 & 0 \\
0 & 0 & \pi J_{33} & \pi J_{34}+\Lambda_{34} \\
0 & 0 & \pi J_{34}-\Lambda_{34} & \pi J_{44}
\end{bmatrix}\,,
\end{equation}
which gives a vanishing spectral gap e.g. if $\Lambda_{12}=0$ and $J_{12}=J_{11}$, consistent with the analysis of the previous section.

Motivated by the simple case discussed so far, we want to study the spectral gap of the model in a more generic scenario.
In general we can anticipate that the spectral gap will be determined (through the matrix $Z$) by the spectral properties of the reservoir, in particular its dimensionality, and the distance between the Majorana zero modes.

To see this, we consider the case in which the Majorana zero modes are tunnel-coupled to a free electron gas in $d$ dimensions, with $d=1,2,3$.
The corresponding Hamiltonian reads
\begin{equation}
H = \sum_{{\mathbf k}} \zeta_{\mathbf k} c^{\dagger}_{\mathbf k}c_{\mathbf k} + \sum_{i=1}^{2N} \left[w_i \gamma_i c^\dagger(\mathbf{r}_i)+\textrm{h.c.}\right]\,.
\end{equation}
The first term is the kinetic energy of the electron gas, with $\zeta_{\mathbf k} = k^2/2m-\mu$, $\mu$ the chemical potential and $\mathbf{k}$ the wave vector. The second term is the coupling between the electron gas and the Majorana zero modes. We adopt a simple point contact model and assume that each Majorana zero mode is coupled locally to the reservoir at a position $\mathbf{r}_i$, with $w_i$ the complex tunneling amplitude. As a further simplifcation, we assume that all the Majorana zero modes are coupled equally strongly to the reservoir, so that the tunneling amplitudes only differ by their phase: $w_i = w e^{i\delta_i}$.

Starting from this concrete model, it is a standard exercise to compute the couplings that enter the Lindblad equation, see App.~\ref{app:toymodels}. We find
\begin{subequations}\label{eq:Js_and_Lambdas}
\begin{equation}
\pi J_{ij}=\Gamma_0\,\cos(\delta_i-\delta_j)
\times\,\begin{cases}
    \dfrac{\sin(k_Fr_{ij})}{k_Fr_{ij}} & d=3 \\
    \\
    J_0(k_Fr_{ij})  & d=2 \\
    \\
    \cos(k_Fr_{ij}) & d=1
\end{cases}
\end{equation}
and
\begin{equation}
\Lambda_{ij} = \Gamma_0\,\sin(\delta_i - \delta_j)\,\times\,\begin{cases}
    \dfrac{\cos(k_Fr_{ij})}{k_Fr_{ij}} & d=3 \\
    \\
    -Y_0(k_Fr_{ij})  & d=2 \\
    \\
    -\sin(k_Fr_{ij}) & d=1
\end{cases}
\end{equation}
\end{subequations}
Here, $\Gamma_0=\pi \nu w^2$ is the tunneling rate at which electrons are exchanged between the reservoir and each Majorana zero mode, with $\nu$ the density of states at the Fermi level; $k_F$ is the Fermi wave vector; $r_{ij}=\abs{\mathbf{r}_i-\mathbf{r}_j}$ is the distance between $i$-th and $j$-th Majorana zero modes; and finally $J_0$ and $Y_0$ are the zero-th order Bessel functions of first and second kind.

Once the positions $\mathbf{r}_i$ of the Majorana zero modes and the tunneling phases $\delta_i$ are set, the equations above can be used to deterministically compute the matrix $Z$ and its spectrum.
With all the Majorana zero modes connected to the same reservoir, the resulting matrix $Z$ is dense.
In particular, the diagonal terms of $Z$ are all equal to $\Gamma_0$, while the off-diagonal terms decay with a power law of the distance between pairs of Majorana zero modes: 
\begin{equation}\label{eq:JandLambda_asymptotes}
\Lambda_{ij}\sim J_{ij} \sim r_{ij}^{(1-d)/2}
\end{equation}
when $k_F r_{ij} \gg 1$\,.

In Fig.~\ref{fig:spectra}, we display the results of a numerical experiment aimed at studying the spectrum of the Liouvillian for reservoirs of different dimensionality. To generate the figure, we have assigned a random position to ten Majorana zero modes using a uniform probability distribution in $[0, L]^d$. The parameter $L$ is a cut-off that sets the maximum possible distance between Majorana zero modes. The phase $\delta_i$ of the tunneling amplitude between each Majorana zero mode and the reservoir was also chosen randomly with a uniform distribution in $[0, 2\pi)$. After this random sampling, we compiled the $10\times 10$ matrix $Z$ using Eq.~\eqref{eq:Js_and_Lambdas}, and found its complex eigenvalues numerically.
Each panel in Fig.~\ref{fig:spectra} shows the position in the complex plane of all the eigenvalues resulting from the diagonalization of five hundred samples.

The top row in the figure shows results for $k_FL=10$, while the bottom row shows results for $k_FL=100$, thus allowing the Majorana zero modes to be further apart from each other.
The observed spread of eigenvalues in the complex plane is the effect of coupling between Majorana zero modes induced by the reservoir: in the absence of off-diagonal elements of $Z$, the eigenvalues would all be equal to $\Gamma_0$.
In $d=2$ and $d=3$, we see that, as $k_FL$ increases, the eigenvalues concentrate closer to $\Gamma_0$, reflecting the decrease in magnitude of the off-diagonal matrix elements of $Z$.
Indeed, in the limit $k_FL \to \infty$ one recovers the limit of disconnected reservoirs.
In $d=1$, on the other hand, the off-diagonal elements of $Z$ do not decay with distance, and the distribution of eigenvalues does not change drastically with $k_FL$.

It is possible to notice other features beyond the dependence on the distance between Majorana zero modes. We recall that the eigenvalues of $Z$ are either purely real or complex-conjugate pairs with opposite imaginary parts. This symmetry is visible in all panels, with a notable tendency, especially in $d=1$ and $d=2$, for part of the spectrum to stick to the real axis. In $d=3$, the eigenvalues which deviates more from $\Gamma_0$ are either on the real axis or on the shifted imaginary axis $\Gamma_0 + i y$, the same behavior captured by the toy model of Sec.~\ref{app:toymodels}. A similar tendency is visible in $d=2$. In $d=1$, instead, there is a characteristic arc-shaped clustering of eigenvalues close the origin in the complex plane.

Overall, the distributions of eigenvalues observed in Fig.~\ref{fig:spectra} reveal some similarities with those of random quadratic Liouvillians of free fermionic systems~\cite{costa2023}, especially for $d=2$ and $d=3$. In our model, randomness is introduced through the position of the Majorana zero modes and the phase of their coupling to the reservoir, while the reservoir itself is a clean system. Due to this combination, unlike in a typical random matrix model, the resulting matrix elements of $Z$ are correlated. We leave to further work a more detailed analysis of the distribution of eigenvalues and their complex level spacings, and of possible similarities and differences with random matrix models.

\begin{figure}[t!]
\includegraphics{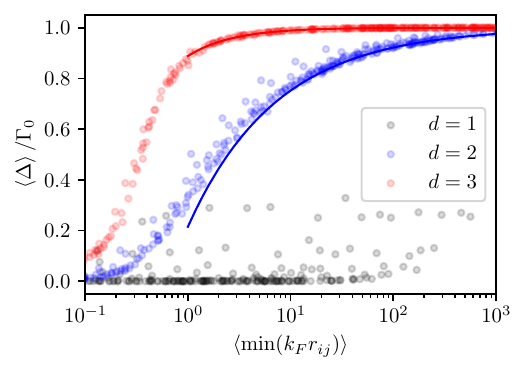}
    \caption{Scaling of the spectral gap with the distance between Majorana zero modes. The plot is generated by computing the spectral gap $\Delta$ and the minimum distance between two Majorana zero modes for one hundred samples with $2N$ Majorana zero modes randomly placed in $[0, L]^d$. We then compute the average gap and average minimum distance over this sample, resulting in a single data point in the scatter plot. We vary $N$ between $2$ and $20$ and $k_FL$ between $1$ and $10^4$ to generate the entire plot.}
    \label{fig:gap_scaling}
\end{figure}

Because the off-diagonal matrix elements decay with the distance between Majorana zero modes in $d=2$ and $d=3$, it is natural to conjecture that the spectral gap $\Delta$ of the Liouvillian is controlled by the minimum distance between two Majorana zero modes, $\min (k_F r_{ij})$. This conjecture is numerically confirmed in Fig.~\ref{fig:gap_scaling}, which aggregates average values of $\Delta$ and $\min(k_F r_{ij})$ from samples with varying number of Majorana zero modes and varying values of the cutoff $L$. The resulting scatter plot shows a clear correlation between the gap and the minimum distance. Even more, the observed curves follow closely the scaling suggested by Eq.~\eqref{eq:JandLambda_asymptotes}, namely
\begin{equation}
1 - \frac{\Delta}{\Gamma_0} \sim [\min(k_Fr_{ij})]^{(1-d)/2}\,.
\end{equation}
This confirms that the decrease in the spectral gap due to the coupling of multiple Majorana zero modes via a single reservoir -- a decrease which causes the slowdown of parity decay discussed in the previous section -- vanishes as a power-law of the distance between Majorana zero modes.

\section{Comparison to numerics}
\label{sec:numerics}

In this section, we compare results obtained from numerical integration of the von Neumann equation, including explicitly the reservoir, to the analytical ones derived from the Lindblad equation in Sec.~\ref{sec:decay}.
The intent of this exercise is to corroborate the validity of the Lindblad approach and probing possible non-Markovian effects.

\begin{figure}
\includegraphics{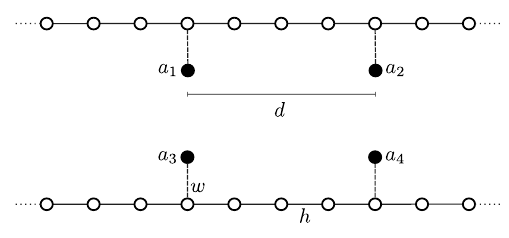}
\caption{Illustration of the model used in the numerical comparison. Majorana zero modes $a_1$ and $a_2$ are coupled via a hopping with strength $w$ to a reservoir given by a chain of Majorana zero modes with uniform nearest neighbor hopping $h$ (a Kitaev chain tuned to its critical point); $a_1$ and $a_2$ are connected to the reservoir $d$ lattice spacings apart. Majorana zero modes $a_3$ and $a_4$ are connected to an identical copy of the same reservoir. Periodic boundary conditions are used for the reservoirs.}
\label{fig:numerics_setup}
\end{figure}

As a model for the reservoir, we choose a Kitaev chain tuned to the critical point, namely a chain of Majorana sites with a uniform hopping (see Fig.~\ref{fig:numerics_setup}).
The Hamiltonian for the model is:
\begin{equation}
\label{eq:CKC}
H = ih\,\sum_{n=1}^{2N} b_n b_{n+1} + iw\,\sum_{i=1}^{2M} a_ib_{n_i}\,.
\end{equation}
Here, $h$ is the hopping along the Kitaev chain, while $w$ is the coupling between the Majorana zero modes and the chain, which we choose to be the same for all Majorana zero modes $a_i$. The index $n_i$ denotes the position along the Kitaev chain at which $a_i$ is coupled. Periodic boundary conditions are assumed at the end of the Kitaev chain.

We have chosen the critical Kitaev chain as the model to be used in the numerical simulations in virtue of its simplicity: it has a single parameter, it is straightforward to implement numerically, and it is analytically solvable.

The critical Kitaev chain has a gapless spectrum, with excitation energies $\zeta(k) = 2h \abs{\sin(k/2)}$.
The couplings $\Lambda_{ij}$ and $J_{ij}$ can be computed analytically from a diagonalization of the critical Kitaev chain, as done in App.~\ref{app:toymodels}.
Letting $d_{ij}$ be the distance (in units of the lattice spacing) between the two sites of the chain to which $a_i$ and $a_j$ are coupled to, the results are
\begin{equation}\label{eq:JLambdaCKC}
J_{ij}=\begin{cases} \dfrac{w^2}{2\pi h} & \textrm{if}\;d_{ij} \;\textrm{is odd,}\\
\\
0 & \textrm{if}\; d_{ij}\;\textrm{is even.} 
\end{cases}
\end{equation}
and
\begin{equation}
\Lambda_{ij}=\begin{cases} 0 & \textrm{if}\;d_{ij}\;\textrm{is odd,} \\
\\
-\dfrac{w^2}{2h} &  \textrm{if}\;d_{ij}\;\textrm{is even.}
\end{cases}
\end{equation}
Note that the couplings are independent of the magnitude of the distance $d_{ij}$, similar to the case of a one-dimensional electron gas in Sec.~\ref{sec:spectrum}, and only depend on its parity.
This curious alternation is due to the fact that it takes two sites of the critical Kitaev chain to define a fermionic mode.
For our purposes it is a very convenient artifact since, by choosing the sites at which Majorana zero modes couple to the chain, it allows us to perform simulations which isolate the roles of $J_{ij}$ and $\Lambda_{ij}$ in the system dynamics.

\subsection{Setup and methods}

We set up a numerical simulation to reproduce the predictions of Fig.~\ref{fig:pairwise} for the case of four Majorana zero modes pairwise coupled to two separate reservoirs.
We therefore set up with two copies of the critical Kitaev chain, with periodic boundary conditions.
The two reservoirs are coupled pairwise to a system of four Majorana zero modes $a_1, a_2, a_3, a_4$: the first reservoir is coupled to $a_1$ and $a_2$ while the second one to $a_3$ and $a_4$.
In both cases the Majorana zero modes are coupled to separate sites of the chain at a distance $d$ from each other, the same for both chains.
Setup this way, the system is symmetric under the exchange of the two reservoirs, and all Majorana zero modes are coupled to a reservoir with equal strengths.

The initial state of the full system is a Gaussian state chosen as follows. The reservoirs are prepared in a thermal state with temperature $T$, namely the state $M_\textrm{th}$ such that
\begin{equation}
OM_\textrm{th}O^T = \bigoplus_{k=1}^N \begin{bmatrix} 0 & \tanh(\beta\zeta_k/2) \\
-\tanh(\beta\zeta_k/2) & 0 \end{bmatrix}
\end{equation}
where $O$, determined numerically, is the orthogonal transformation that ``diagonalizes'' the real-space reservoir Hamiltonian, as in Eq.~\eqref{eq:standard_form_bath_hamiltonian}.

The four Majorana zero modes $a_1, a_2, a_3, a_4$ are prepared in one of the two initial Gaussian states $M^X$ and $M^Z$ already defined in Sec.~\ref{sec:decay}, see Eqs.~\ref{eq:MX_MZ}. 
The covariance matrix of the full system is then a direct sum, either $M(0) = M_\textrm{th}\oplus M^X$ or $M(0)=M_\textrm{th}\oplus M^Z$, without any initial correlation between the Majorana zero modes and the reservoirs.

\begin{figure*}[ht]
\includegraphics{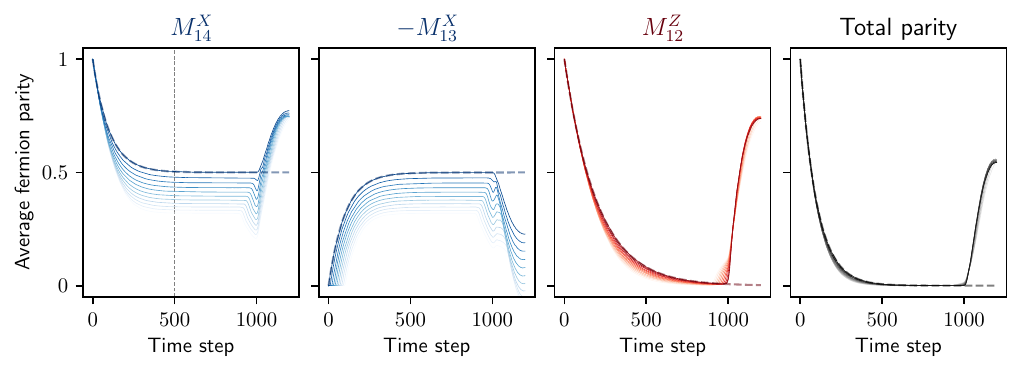}
\caption{Time evolution of various parity expectation values for the four-Majorana, two-reservoir system of Fig.~\ref{fig:numerics_setup}, for even distances $d$ between Majorana zero modes. The blue and red curves correspond to X-type and Z-type initial conditions respectively. The rightmost panel shows the time evolution of the total parity, which is the same for both initial conditions. The dashed curves are the Lindblad predictions, which for this model are independent of $d$, while the solid lines represent numerical results for different values of $d$, increasing in steps of 10 from $d=0$ to $d=100$ from the darker to the lighter lines. The dashed vertical line in the leftmost panel indicates the time slice at which the further results presented in Fig.~\ref{fig:plateau_scaling} are computed.}
\label{fig:num_even}
\end{figure*}

\begin{figure}
\includegraphics{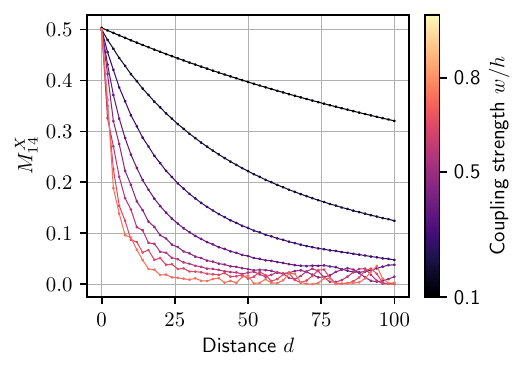}
\caption{Evolution of the parity expectation value $M_{14}^X$ computed at $t=500$ as a function of the distance $d$ between the Majorana zero modes. Different curves correspond to increasing values of the coupling strength $w$, from $0.1$ to $0.8$ in steps of $0.1$. The plot shows more systematically the decrease in the expectation value, computed at the plateau reached in Fig.~\ref{fig:num_even} before the onset of the finite-size revival, when increasing distance or coupling strength.}
\label{fig:plateau_scaling}
\end{figure}

As we have seen before, in the Majorana basis, the Hamiltonian of the full system under investigation can be encoded into a real antisymmetric matrix $A$. The time evolution of the full covariance matrix is then dictated by the von Neumann equation,
\begin{equation}
\label{eq:von Neumann}
\dot{M} = \frac{1}{2}[A, M]\,.
\end{equation}
The solution is
\begin{equation}
M(t) = U(t) M(0) U^{-1}(t)\,,\quad U(t) = e^{At/2}\,.
\end{equation}
We divide the time evolution in time steps of length $dt$ and we iterate this solution starting from $M(0)$.
Since $A$ is independent of time, we need to compute $U(dt)$ only once, at the beginning of the simulation.
After each iteration, we enforce the antisymmetry $M=-M^T$ in order to remove any spurious symmetric part that may result from the numerical matrix multiplication and so mitigate the accumulation of floating point errors.

\subsection{Numerical results}

We carried out the simulation as described in the previous section, with two Kitaev chains each long a thousand sites ($N=500$). 
We carried simulations for both even and odd distances $d$ between the Majorana zero modes, starting from $d=0$ and $1$ and increasing $d$ in steps of ten until $d=100$ and $101$.
In terms of parameters, we set $h=1$ and $w=0.1$.
Unless noted otherwise, the temperature of the reservoir is set to $T=0$.
Finally, we used a time step $dt=1$ and carried out the simulation for 1200 time steps.

For this simulation, the analytical predictions coming from Sec.~\ref{sec:decay} are those of Eq.~\eqref{eq:Ztype_evolution} and Eq.~\eqref{eq:Xtype_evolution} replacing the values of the couplings given in Eq.~\eqref{eq:JLambdaCKC}. For Z-type initial conditions, we have
\begin{equation}
M^Z(t) = e^{-w^2t/2h}\,M^Z(0)\,,
\end{equation}
regardless of the distance $d$ between Majorana zero modes. For $X$-type initial conditions, the time-evolution depends on the parity of $d$. For even values of $d$, $\Lambda_{ij}=0$ and we obtain
\begin{align}
M^X_{13}(t)&=-e^{-w^2/2ht}\sinh\left(\frac{w^2t}{2h}\right)\,,\\
M^X_{14}(t)&=e^{-w^2/2ht}\cosh\left(\frac{w^2t}{2h}\right)\,.
\end{align}

In Figure~\ref{fig:num_even}, we compare the corresponding numerical results (solid lines) to these predictions (dashed lines).
For the sake of simplicity, we only show some select matrix elements of the covariance matrix: results for $X$-type initial conditions are shown in blue while results for $Z$-type initial conditions are shown in red.
The evolution of the total parity is the same for both initial conditions and is shown in black. 
The different solid lines correspond to increasing distances between the Majorana zero modes, with larger distances gradually fading towards a white color.

Qualitatively, the numerical results reproduce the trends expected from the Lindblad equation.
In particular, they confirm what is perhaps the most interesting feature, the saturation of $M^X_{14}$ and $M^X_{13}$ to $\pm 1/2$ due to the presence of a second steady state.
More quantitatively, we find that the results of the numerics approach the Lindblad predictions better for progressively smaller distances between the Majoranas.
This reflects the finite time, growing linearly with distance in the simulation, required to establish correlation between distant Majorana zero modes.
Furthermore, the numerics show a visible revival of expectation values as the time approaches $\approx 1000$ time steps.
Note that, for the parameters chosen in the simulation, the Fermi velocity in the reservoir is unity, thus this is the time scale required to probe the finite size of the reservoir.

\begin{figure*}
\includegraphics{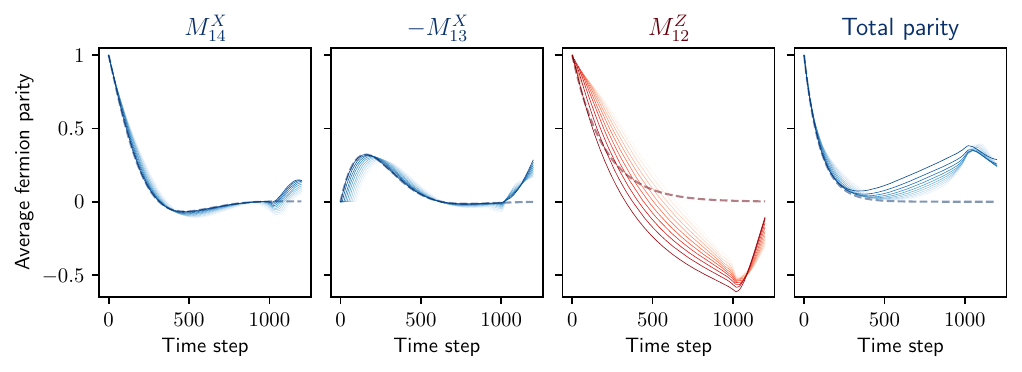}
\caption{Results for the same time evolution protocol as in Fig.~\ref{fig:num_even}, but for odd distances $d$ between Majorana zero modes: in this case, $d$ increases in steps of 10 from $d=1$ to $d=101$ from the darker to the lighter lines.}
\label{fig:num_odd}
\end{figure*}

These two features of the full numerical solution do not appear in the Lindblad solutions because they are washed away by the Markovian assumption discussed in Sec.~\ref{app:lindblad}.
Both effects would disappear in the limit of a reservoir of infinite size and bandwidth, and thus they can be classified as non-Markovian effects.
While the revival is essentially an artifact of our simulations, which are limited by system size, the time delay effect could be relevant in experimental situations with distant Majorana zero modes and reservoirs with a small Fermi velocity.
In particular, if the propagation time between two Majorana zero modes via the reservoir becomes longer than the inverse tunneling rate between a Majorana zero mode and the reservoir, then the connectedness of the reservoir stops playing a role.

This effect is illustrated in Fig.~\ref{fig:plateau_scaling}, where we show the evolution of the plateau reached in Fig.~\ref{fig:num_even} before the onset of finite-size effects.
When increasing the distance between Majorana zero modes or the coupling strength between the Majorana zero modes and the reservoir, the numerical calculations deviate ever more from the steady-state value predicted by the Lindblad with non-local terms.
Indeed, when correlations between distant zero modes cannot be established quickly enough, the non-local effects vanish and the predictions for a disconnected reservoir are re-established.

In Figure~\ref{fig:num_odd}, we show results corresponding to odd distances $d$ between the Majorana zero modes. For odd $d$, the Lindblad prediction replaces hyperbolic functions in the time evolution with trigonometric ones (this is the case with discriminant $\Delta_{12}$ purely imaginary in Sec.~\ref{sec:decay}):
\begin{align}
M^X_{13}(t)&=-e^{-w^2/2ht}\sin\left(\frac{w^2t}{2h}\right)\,,\\
M^X_{14}(t)&=e^{-w^2/2ht}\cos\left(\frac{w^2t}{2h}\right)\,.
\end{align}
The comparison again confirms the essential aspects of the Lindblad prediction, showing a damped oscillation of the expectation values for $X$-type initial conditions.
The same non-Markovian effects discussed in regard with Fig.~\ref{fig:num_even} are also visible in Fig.~\ref{fig:num_odd}.

A much larger deviation from the analytical predictions can be see in the time evolution of $M_{12}$ and of the total parity: we show explicitly the disagreement for $M_{12}^Z$ and for the total parity in the X-type case, however a qualitatively similar disagreement for any expectation value involving the products $a_1 a_2$ or $a_3 a_4$.
Thus, the mismatch can be traced back to the presence of the Lamb shifts terms $\Lambda_{ij}$, which in our reservoir model only occur for odd distances.

\begin{figure}
\includegraphics{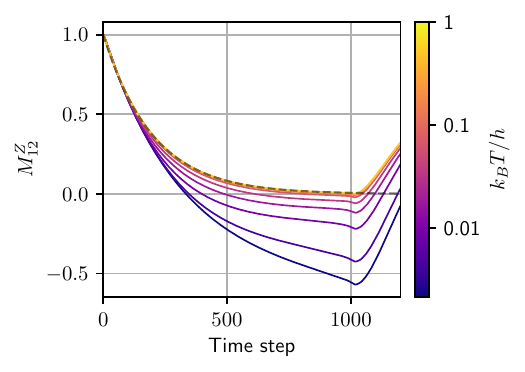}
\caption{Time evolution of the expectation value of $-2i a_1 a_2$ shown for different reservoir temperatures, for Z-type initial conditions and a distance $d=1$. The numerical curves (solid lines) gradually approach the analytical result obtained from the Lindblad equation (dashed line) as the temperature increases.}
\label{fig:num_finite_t}
\end{figure}

It is indeed known that the Lindblad equation generically evolves the system towards a thermal state with Boltzmann weights that do not include the Lamb shift of energy levels~\cite{mori2008}.
To reconcile this issue, we investigated the behavior of expectation values as a function of the temperature of the reservoir, as shown in Figure~\ref{fig:num_finite_t}.
We see that the time evolution gradually approaches the Lindblad prediction as the temperature of the reservoir increases.

We interpret these results as follows.
When the temperature of the reservoir becomes larger than the Lamb shifts, we recover the fully mixed steady state corresponding to an exact degeneracy of Majorana zero modes, as predicted by the Lindblad equation.
On the other hand, if the temperature of the reservoir is smaller than the Lamb shift, the steady state (which is not fully accessible in our simulation due to finite size-effects) deviates from the Lindblad prediction.

\section{Conclusions}
\label{sec:conclusions}

In this work, we have investigated the decoherence of Majorana zero modes coupled to a gapless fermionic reservoir. We have derived a Lindblad equation valid for any non-interacting reservoir Hamiltonian. We have shown that, if a single reservoir couples to multiple Majorana zero modes, there appear non-local jump operators which can lead to a slowdown of the decoherence compared to the case of disconnected reservoirs. We have analyzed the spectral properties of the Lindblad equation for a simple, free-electron reservoir and showed a power-law dependence of the spectral gap on the distance between Majorana zero modes. An unbiased numerical solution of the von Neumann equation revealed that the non-local couplings take a finite time to establish, but otherwise confirms the qualitative predictions of the Lindblad model.

Our results do not alter in any way the criterion that the presence of fermionic reservoirs has to be avoided in Majorana qubit devices.
However, this criterion may not always be applicable in practice, especially in intermediate experiments testing new materials and device designs, where transport measurements remain essential.
Furthermore, our results indicate that the presence of a fermionic reservoir leads to a dissipative dynamics which is interesting in its own regard.

Although our modeling did not include the fermion parity measurement apparatus explicitly, we emphasize that the protocol described in Sec.~\ref{sec:decay} to isolate the non-local decoherence effect could be experimentally implemented e.g. using tetron devices~\cite{karzig2017,microsoft2025}.
In this setting, a two-dimensional electron gas could be used as an intentional fermionic reservoir with tunable couplings, in order to test the dissipative dynamics of Majorana zero modes.
Conceptually, the presence of two separate baths is not strictly necessary: it would be sufficient to couple two out of four Majorana zero modes to a common reservoir.

Our findings leave some immediate open questions. Can the Lindblad equation be refined, perhaps with a better time coarse-graining, to capture the time delay occurring before non-local couplings are established? Can the spectral properties of the Lindblad equation for the free-electron model be described with random-matrix theory? Can the open-system dynamics reveal a difference between unpaired Majorana zero modes and local zero-energy Andreev bound state~\cite{mishmash2020}? We leave the exploration of these questions for further work.

There are also several possible extensions of our work. It would be important to include disorder in the reservoir and interactions, both between the Majorana zero modes and in the reservoir.
Interactions between the Majorana zero modes due to charging energy of the topological superconductor, in particular, are important to model Majorana-based topological qubits~\cite{karzig2017} and predicted transport phenomena such as the electron teleportation~\cite{fu2010} and the topological Kondo effect~\cite{beri2012}.
It would also be interesting to change the perspective on the model of Eq.~\eqref{eq:microscopic_model} and study transport through a reservoir coupled to many Majorana zero modes.
Finally, the Lindblad equation for Majorana zero modes developed here could be used to study non-Abelian braiding~\cite{beenakker2020} in the presence of dissipation, for instance in the case of vortices weakly coupled to edge states.

\subsection*{Data availability}

All data generated and code used in this work are available on Zenodo~\cite{zenodo}.

\begin{acknowledgments}
We thank Tomaž Prosen for a useful discussion as well as Mario Alessandrini, Sergio Caprara, and Giulia Venditti for discussions and collaborations on related topics.
Likewise, we thank Michele Burrello and Bela Bauer for feedback on the manuscript and Bela Bauer in particular for alerting us to Ref.~\cite{campbell2015}.
M.G. and S.B. acknowledge financial support from the PNRR MUR project PE0000023-NQSTI, and specifically the project ‘Topological Phases of Matter, Superconductivity, and Heterostructures’ Partenariato Esteso 4-Spoke 5 (n. PE4221852A63A88D), and from the ‘Ateneo Research Projects’ of the Sapienza University of Rome: ‘Equilibrium and out-of-equilibrium properties of low-dimensional disordered and inhomogeneous superconductors’ (n. RM12017 2A8CC7CC7), ‘Competing phases and non-equilibrium phenomena in low-dimensional systems with microscopic disorder and nanoscale inhomogeneities’ (n. RM12117A4A7FD11B).
\end{acknowledgments}

\appendix

\section{Derivation of the Lindblad equation}
\label{app:lindblad}

Our starting point is the microscopic Hamiltonian~\eqref{eq:microscopic_model}, which we can rewrite here as
\begin{equation}
H = \frac{i}{2} a^T W b + \frac{i}{2} b^T B b \equiv H_{AB}+H_B\,,
\end{equation}
where $a=(a_1, \dots, a_{2N})$ gathers all the Majorana operators of our system and $b=(b_1, \dots, b_{2K})$ gathers all the Majorana operators for the reservoir.
We recall the anti-commutators $\{{a}_{i},{a}_{j}\}=\delta_{ij}$, $\{b_n,b_m\}=\delta_{nm}$ and  $\{a_i,b_n\}=0$.
Formally, we assume that the reservoir hosts $K$ fermionic modes, with $K\gg N$; in fact we will later take the limit $K\to \infty$ when integrating out the reservoir.
Note that $W$ is a rectangular matrix of size $2N \times 2K$, whose elements are the tunnel couplings $W_{in}$.

The $2K\times 2K$ matrix $B$ is real and antisymmetric: $B=-B^T$. As such, there exists an orthogonal matrix $O=O^T$ that brings into the standard form
\begin{equation}\label{eq:standard_form_bath_hamiltonian}
O B O^T = \bigoplus_{k=1}^K \begin{bmatrix}
    0 & \zeta_k\\ -\zeta_k & 0
\end{bmatrix}\,,
\end{equation}
Thus, the Hamiltonian can be written as
\begin{equation}
H = \frac{i}{2} a^T \bar{W}\bar{b} + i\sum_{k=1}^K \zeta_k\, \bar{b}_{2k-1}\bar{b}_{2k}
\end{equation}
with $\bar{b}=O b$ and $\bar{W}=WO^T$. The reservoir Hamiltonian $H_B$ is now a sum of commuting terms, each corresponding to a single-particle eigenmode which can be empty ($i\bar{b}_{2k-1}\bar{b}_{2k}=-\tfrac{1}{2}$) or occupied ($i\bar{b}_{2k-1}\bar{b}_{2k}=+\tfrac{1}{2}$).

To describe the time evolution of the full system we adopt the interaction picture with respect to $H_B$.
In this picture, the Majorana operators $a_i$ do not evolve with time since $[a_i, H_B]=0$.
On the other hand, the reservoir operators do:
\begin{equation}
\bar{b}(t) = \Theta (t) \bar{b}(0)\,,
\end{equation}
with
\begin{equation}
\quad \Theta(t)=\bigoplus_{k=1}^K \begin{bmatrix}
    \cos(\zeta_kt) &\sin(\zeta_kt) \\ -\sin(\zeta_kt) & \cos(\zeta_kt)
\end{bmatrix}\,.
\end{equation}
As a consequence, the tunneling Hamiltonian also evolves with time:
\begin{equation}
H_{AB}(t) =\frac{i}{2}\sum_{i=1}^{2N}\,a_iB_i(t)\,,
\end{equation}
where we introduced channel operators $B_i(t)$,
\begin{equation}
\quad B_i(t) = \sum_{n=1}^{2K} \bar{W}_{in}\bar{b}_n(t)\,.
\end{equation}

In the interaction picture, the time evolution of the total density matrix $\rho_\textrm{tot}$ describing both the system and reservoir is given by the Liouville-von Neumann equation:
\begin{equation}
\dot{\rho}_\textrm{tot} = -i\,[{H}_{AB}(t), \rho_\textrm{tot}(t)].
\end{equation}
In integral form, and introducing an initial condition $\rho_\textrm{tot}(0)$, one has
\begin{equation}
\rho_\textrm{tot}(t) = \rho_\textrm{tot}(0) + \frac{1}{2}\sum_{i=1}^{2N}\,\int_0^t ds\,[a_i{B}_i(s),\rho_\textrm{tot}(s)] . 
\end{equation}
Replacing this equation into the Liouville-von Neumann equation, we obtain
\begin{align}\label{eq:LvN_with_boundary_term}\nonumber
\dot{\rho}_\textrm{tot} &= \frac{1}{2} \sum_{i=1}^{2N} [a_iB_i(t),\rho_\textrm{tot}(0)] \\ 
&+ \frac{1}{4} \sum_{i,j=1}^{2N} \int_0^t ds\, [a_iB_i(t),[a_j{B}_j(s),\rho_\textrm{tot}(s)]] .
\end{align}
To proceed further, we assume that the initial state is separable:
\begin{equation}\label{eq:initial_condition}
\rho_\textrm{tot}(0) = \rho(0)\,\rho_B\,,
\end{equation}
with $\rho(0)$ a reduced density matrix for the Majorana zero modes at time $t=0$, and ${\rho}_B$ a thermal density matrix for the reservoir,
\begin{align}\nonumber
{\rho}_B &= Z^{-1}\,\exp(-\beta H_B) \\ 
&= Z^{-1} \exp\left(-i\beta \sum_{k=1}^K \zeta_k \bar{b}_{2k-1}\bar{b}_{2k}\right)\,.
\end{align}
with $Z=\tr_B[\exp(-\beta{H}_B)]$ and $\beta=1/T$. Note that $[a_i, \rho_B]=0$ and thus $[\rho(0), \rho_B]=0$.

The density matrix of the Majorana zero modes at time~$t$ can be obtained via the partial trace over the degrees of freedom of the reservoir:
\begin{equation}
\rho(t) = \tr_B\,[\rho_\textrm{tot}(t)] .
\end{equation}
Taking the partial trace over both sides of Eq.~\eqref{eq:LvN_with_boundary_term}, we obtain
\begin{equation}
\dot \rho = \frac{1}{4}\sum_{i,j=1}^{2N} \int_0^t ds\,\tr_B [a_iB_i(t),[a_j{B}_j(s),\rho_\textrm{tot}(s)]]\, .
\end{equation}
Note that the boundary term in Eq.~\eqref{eq:LvN_with_boundary_term} vanishes upon taking the partial trace, since $B_i(t)$ is linear in the reservoir Majorana operator, and thus $\tr_B\left[\rho_B B_i(t)\right] = 0$.

So far, no approximations have been made. If the tunneling between the reservoir and the Majorana zero modes is weak, it is sensible to consider the effect of the Majorana zero modes on the state of the reservoir to be negligible.
That is, assume that the density matrix remains separable at all times, and that the density matrix of the reservoir remains constant:
\begin{equation}
\rho_\textrm{tot}(t) = \rho(t)\rho_B .
\end{equation}
This is the Born approximation. Inserting this ansatz into the integrand, we may take $\rho(t)$ outside of the partial trace, noting that
\begin{equation}
[{B}_j(s){B}_i(t),\rho(t)] = 0\,,
\end{equation}
since the product of two reservoir operators is bi-linear in the Majorana operators of the reservoir and $\rho(t)$ only depends on the Majorana operators $a_i$. 
After some algebra, we obtain a closed integro-differential equation for $\rho(t)$:
\begin{align}\label{eq:master_eq_with_born_approx}\nonumber
\dot \rho = -\frac{1}{4}\sum_{i,j=1}^{2N} \int_0^t ds\,\big\{&C_{ij}(t-s)\,[a_i,a_j\,\rho(s)] \\ + &C_{ji}(s-t)\,[\rho(s)a_j,a_i]\big\},
\end{align}
where we have introduced the correlation function of the reservoir
\begin{equation}
C_{ij}(t-s) = \tr_B\,[{B}_i(t){B}_j(s)\,{\rho}_B]\,.
\end{equation}
Since $[\rho_B, H_B]=0$, the correlation function only depends on the difference $t-s$ of the two time arguments appearing inside the partial trace on the right hand side. 

The master equation with the Born approximation, Eq.~\eqref{eq:master_eq_with_born_approx}, is still formidable to solve because it has a kernel which is non-local in time. 
It is necessary to integrate the entire evolution of ${\sigma}$ from $s=0$ until $s=t$ to predict its evolution from $t$ to $t+dt$.
In other words, the memory of the previous interactions with the reservoir is still preserved.
To simplify the equation further, at this point the \emph{Markov approximation} is made, which involves two logically distinct steps.

In the first step, one imagines that the correlation function $C_{ij}(t)$ decays quickly for both positive and negative arguments.
By ``quickly'', we mean that $C_{ij}(t)$ approaches zero faster than the rate of change of the reduced density matrix ${\rho}$.
This situation can always be realized by decreasing the coupling strength between the Majorana zero modes and the reservoir.
We discuss the validity of this assumption later in Section~\ref{sec:numerics} on the base of numerical simulations, and also in Appendix~\ref{app:reservoir_correlation_functions} from an analytical standpoint.

If $C_{ij}(t)$ decreases quickly when $t$ deviates from zero, the dominant contribution of the integrand in Eq.~\eqref{eq:master_eq_with_born_approx} to the integral comes when $s\approx t$, at the right end of the domain of integration. Then, it is possible to replace $\rho(s)$ with $\rho(t)$ in the integrand:
\begin{align}\label{eq:bloch_redfield}\nonumber
\dot \rho = -\frac{1}{4}\sum_{i,j=1}^{2N} \int_0^t ds \big\{&C_{ij}(t-s)\,[a_i,a_j\,\rho(t)] \\ + &C_{ji}(s-t)\,[\rho(t)a_j,a_i]\big\}.
\end{align}
This is the Bloch-Redfield master equation for our model. Although local in time, it has the inconvenience that, because of the finite domain of integration, the coefficients on the right hand side will be time-dependent. To remove this inconvenience, the second step of the Markov approximation kicks in. Because $C_{ij}(t)$ decreases rapidly away from zero, we can extend the domain of integration from $[0,t]$ to $(-\infty, t]$, since the integrand is vanishingly small in the added range. Then, a change in integration variable from $s$ to $t-s$ bring the master equation to have the form:
\begin{align}\label{eq:born-markov_me}\nonumber
\dot \rho = -\frac{1}{4}\sum_{i,j=1}^{2N}\bigg\{&\left[\int_0^\infty ds\,C_{ij}(s)\right]\,[a_i,a_j\,\rho(t)] \\ +& \left[\int_0^\infty ds\,C_{ji}(-s)\right]\,[\rho(t)\,a_j,a_i]\bigg\}.
\end{align}
This completes the derivation of the master equation under the Born-Markov approximation.
No further approximation is needed to arrive at the Lindblad equation~\eqref{eq:liouvillian}: we only need to compute the correlation functions more explicitly. Using
\begin{equation}
\tr\left(\bar{b}_{2k-1}\bar{b}_{2k}\rho_B\right)= \frac{i}{2}\tanh\left(\tfrac{1}{2}\beta\zeta_k\right)
\end{equation}
and after some algebra, one obtains
\begin{equation}
\label{eq:reservoir correlation}
\begin{split}
C_{ij}(t) &= \tfrac{1}{2}\sum_k\left[\cos(\zeta_kt)S_{ijk}+\sin(\zeta_kt)A_{ijk}\right]\\ &+\tfrac{i}{2}\sum_k\,\tanh\left(\tfrac{1}{2}\beta\zeta_k\right)\,\left[\cos(\zeta_k t)A_{ijk}-\sin(\zeta_kt)S_{ijk}\right]\,,
\end{split}
\end{equation}
with $S_{ijk}$ and $A_{ijk}$ the following combinations of tunneling matrix elements respectively:
\begin{subequations}\label{eqs:S_and_A}
\begin{align}
S_{ijk} &= \bar{W}_{i,2k-1}\bar{W}_{j,2k-1}+\bar{W}_{i,2k}\bar{W}_{j,2k}\,,\\
A_{ijk} &= \bar{W}_{i,2k-1}\bar{W}_{j,2k}-\bar{W}_{i,2k}\bar{W}_{j,2k-1}\,.
\end{align}
\end{subequations}
We note that $S_{ijk}$ is symmetric in the $i,j$ indices, while $A_{ijk}$ is antisymmetric.
Note that these are the coefficients first mentioned in Sec.~\ref{sec:model}, in Eq.~\ref{eq:Lambda_ij_J_ij}.

We may now integrate the reservoir correlation function in time to obtain the form of the coefficients which enter the master equation. We regularize the integrals, introducing an exponential cutoff $e^{-\delta t}$, and sending $\delta\to 0$ after the integration. This results in the following prescription:
\begin{align}\nonumber
&\int_0^\infty \cos(\zeta_k t) dt \mapsto \pi\delta(\zeta_k)\\
&\int_0^\infty \sin(\zeta_k t) dt \mapsto \textrm{p.v.} \frac{1}{\zeta_k}
\end{align}
The principal value should be interpreted as a prescription on how to do the integrals over the reservoir degrees of freedom after taking the continuum limit for the quantum number $k$. From now on we will not indicate the need to take principal value explicitly, it will be given for granted when a factor $\zeta_k^{-1}$ is present in the summand or integrand.
And so we obtain:
\begin{align}\nonumber
\int_0^\infty ds\,C_{ij}(s) &= \pi J_{ij} + \Lambda_{ij} - i Y_{ij}\,,
\end{align}
where the coefficient $J_{ij}$ and $\Lambda_{ij}$ first defined in Eq.~\eqref{eq:Lambda_ij_J_ij} appear, and
\begin{equation}
Y_{ij}=\tfrac{1}{2}\sum_k \tanh\left(\tfrac{1}{2}\beta\zeta_k\right)  \zeta_k^{-1}S_{ijk}\,.
\end{equation}
To arrive at this result, we have used that the product $\delta(\zeta_k)\tanh\left(\tfrac{1}{2}\beta\zeta_k\right)$ is identically zero.
We note that $Y_{ij}$ is purely symmetric, $Y_{ij}=Y_{ji}$. The other integral appearing in the master equation is simply the complex conjugate of the one just computed:
\begin{align}
\int_0^\infty ds\,C_{ji}(-s) = \pi J_{ij} + \Lambda_{ij} + iY_{ij}\,.
\end{align}
Replacing these expressions in the master equation, we arrive at
\begin{equation}
\dot \rho = -\frac{1}{4} \sum_{ij=1}^{2N} (\pi J_{ij}+\Lambda_{ij})\Sigma_{ij} + \frac{i}{4}\sum_{ij=1}^{2N} Y_{ij}\Delta_{ij}\,,
\end{equation}
with
\begin{align}\nonumber
\Sigma_{ij} &=[a_i,a_j\,\rho_A(t)] + [\rho_A(t)a_j,a_i]\,,\\
\Delta_{ij} &=[a_i,a_j\,\rho_A(t)] - [\rho_A(t)a_j,a_i]\,.
\end{align}
We observe that $\Delta_{ij}$ is antisymmetric in the indices $ij$, so the second term is vanishing $\sum_{ij}Y_{ij}\Delta_{ij}=0$. Instead, $\Sigma_{ij}$ contains both a symmetric and an anti-symmetric part, which combine with $J_{ij}$ and $\Lambda_{ij}$ respectively to give non-zero contributions to the right hand side.
Gathering all non-vanishing terms, we obtain Eq.~\eqref{eq:liouvillian}:
\begin{equation}\label{eq:lindblad_final}
\dot \rho = -\frac{1}{4}\sum_{ij} \Lambda_{ij} [a_ia_j,\rho] -\frac{\pi}{4}\sum_i\,J_{ii}\rho + \frac{\pi}{2}\sum_{ij}J_{ij}\,a_i\rho a_j\,.
\end{equation}

Lastly, we prove that the matrix $J$ is semi-positive-definite. Consider an arbitrary $2M$-dimensional vector $v_i$ and compute the inner product $v^T J v$, one has:
\begin{align}\nonumber
v^TJv &= \sum_k \delta(\zeta_k) \sum_{ij} v_i S_{ijk} v_j \\\nonumber &= \sum_k \delta(\zeta_k) [(\sum_i v_i T_{i,2k-1})^2+(\sum_i v_i T_{i,2k})^2] \geq 0 \,.
\end{align}
This property of the matrix $J$ completes the demonstration that the master equation is indeed of the Lindblad form.

Note that no approximation was invoked in going from Eq.~\eqref{eq:born-markov_me} to Eq.~\eqref{eq:lindblad_final}.
Usually this step involves a secular or rotating-wave approximation, even though recent formulations of the Lindblad equation avoid this requirement~\cite{nathan2020,mozgunov2020}.
In our case, the degeneracy of the Majorana zero modes makes the dynamics of the closed system trivial and thus avoid the need of any secular approximation to arrive at the Lindblad equation.

\section{Covariance matrix formalism}
\label{app:covariance_matrix_evolution}

In this section, we provide the derivation of Eq.~\ref{eq:dotM}, which is the dynamical equation obeyed by the covariance matrix for the Lindblad time evolution of a Gaussian state.

We consider that system $H_{A}$ is coupled to an external reservoir $H_{B}$, in the manner described in Eq.~\ref{eq:microscopic_model}. We write an effective Lindblad equation (Eq.~\ref{eq:lindblad}) for the time evolution of the reduced density matrix $\rho$ corresponding to the $A$ system. Using this, we now derive the corresponding equation of motion for the covariance matrix, given that one starts from a Gaussian initial state.

One starts from the definition given in Eq.~\ref{eq:covariance matrix}, writes the expectation value as a trace, and puts the time derivative on $\rho$- 
\begin{align}\nonumber
\dot M_{ij} &= -i\tr([a_i,a_j]\dot{\rho}) \\ 
&= -i\tr\left([a_i,a_j] \mathcal{L}(\rho) \right).
\end{align}

We first consider the unitary part, which upon using the cyclic property of the trace, simplifies to- 
\begin{equation}
\begin{split}
\frac{i}{4} \sum_{kl} \Lambda_{kl}(\av{a_i a_j a_k a_l} - \av{a_j a_i a_k a_l} \\ - \av{a_k a_l a_i a_j} + \av{a_k a_l a_j a_i}) .
\end{split}
\end{equation}
Using Wick's theorem (Eq.~\ref{eq:wick}), we may now decompose the expectation values of four Majorana operators in terms of those containing two operators. The basic identity is
\begin{equation}
\av{a_i a_j a_k a_l} = \av{a_i a_j}\av{a_k a_l} + \av{a_i a_l}\av{a_j a_k} - \av{a_i a_k}\av{a_j a_l} .     
\end{equation}
Moreover, using the anticommutation rules of the Majorana operators, we may rewrite the following products of expectation values as- 
\begin{align} \nonumber
\av{a_k a_j}\av{a_l a_i} &= \delta_{jk}\delta_{il} - \delta_{il}\av{a_j a_k} \\ &- \delta_{jk}\av{a_i a_l} + \av{a_j a_k}\av{a_i a_l} \nonumber \\ 
\av{a_k a_j}\av{a_l a_i} &= \delta_{jk}\delta_{il} - \delta_{il}\av{a_j a_k} \nonumber \\ &- \delta_{jk}\av{a_i a_l} + \av{a_j a_k}\av{a_i a_l} .
\end{align}
Using the above identities, the unitary part reduces to the following expression.
\begin{equation}
\begin{split}
\frac{i}{2}\sum_{kl}\Lambda_{kl} (\delta_{il}\av{a_j a_k} &+ \delta_{jk}\av{a_i a_l} - \delta_{ik}\av{a_j a_l} \\ &- \delta_{lj}\av{a_i a_k} - \delta_{jk}\delta_{il} + \delta_{ki}\delta_{lj}) . 
\end{split}
\end{equation}
We now write down the expansion of $[\Lambda,M]_{ij}$ as- 
\begin{align}\nonumber
[\Lambda,M]_{ij} &= \sum_{k}(\Lambda_{ik}M_{kj}-M_{ik}\Lambda_{kj}) \\ \nonumber &= -i\sum_{k}\Lambda_{ik}(\av{a_k a_j} - \av{a_j a_k}) \\ &+ \sum_{k}(\av{a_i a_k} - \av{a_k a_i})\Lambda_{kj} . 
\end{align}
Upon comparing this with the preceding expression for the unitary part, considering $\Lambda_{ji}=-\Lambda_{ij}$, and after doing some algebra, we infer that
\begin{equation}
\dot M \vert_{unitary} = \frac{1}{2} [\Lambda,M] .    
\end{equation}

Next, we consider the dissipative parts of the Lindblad equation. The first `diagonal' part just involves the trace of the $J$ matrix, which yields a simple expression- 
\begin{equation}
i\frac{\pi}{4}\sum_k (\av{a_i a_j} - \av{a_j a_i}) .    
\end{equation}
The other `off-diagonal' part can be rewritten as
\begin{equation}
-i\frac{\pi}{2} \sum_{kl} J_{kl} (\av{a_i a_j a_k a_l} - \av{a_l a_j a_i a_k}) ,
\end{equation}
again using the cyclicity of the trace. This term can then be factorized using Wick's theorem as previously. The result is as follows.
\begin{equation}
\begin{split}
-i\frac{\pi}{2} \sum_{kl} J_{kl} (2\av{a_l a_i}\av{a_j a_k} - 2\av{a_l a_j}\av{a_i a_k} \\ + \av{a_l a_k}(\av{a_i a_{j}}-\av{a_j a_i}) .    
\end{split}
\end{equation}
Next, we use the property $J_{kl}=J_{lk}$ to simplify the above expression as follows.
\begin{equation}
\begin{split}
-i\frac{\pi}{2}(&\sum_{kl} J_{kl} \av{a_l a_k} (\av{a_i a_j} - \av{a_j a_i}) -2 J_{ji} \\ + 2&\sum_{k} \av{a_j a_k} + 2\sum_{k} J_{kj} \av{a_k a_i}) .    
\end{split}
\end{equation}
Again, due to the anticommutation relations- 
\begin{align}\nonumber
\sum_{k} J_{ik} \av{a_j a_k} &= J_{ij} - \sum_{k} J_{ij} \av{a_k a_j} \\ \sum_{k} J_{kj} \av{a_k a_i} &= J_{ij} - \sum_{k} J_{kj} \av{a_i a_k} .    
\end{align}
Hence, we note that in the top line of the last expression, the first piece cancels with the contribution from the `diagonal' part, while the term $-2J_{ji}$ is cancelled by the two $J_{ij}$ terms generated by the previous anti-commutator.
Now, we consider the following identity.
\begin{align}\nonumber
\{J,M\}_{ij} &= \sum_{k} (J_{ik}M_{kj} + M_{ik}J_{kj}) \\ \nonumber 
&= -i\sum_{k} J_{ik} (\av{a_k a_j} - \av{a_j a_k}) \\ &-i \sum_{k} (\av{a_i a_k} - \av{a_k a_i})J_{kj} .
\end{align}
So, upon comparing terms, we realize that- 
\begin{equation}
\dot M \vert_{dissipative} = -\frac{\pi}{2} \{J, M\} .    
\end{equation}
By combining this with the result for the unitary part, we finally get- 
\begin{align}\nonumber
\dot M &= \dot M \vert_{unitary} + \dot M \vert_{dissipative} \\ &=\frac{1}{2} [\Lambda, M] - \frac{\pi}{2} \{J, M\} .
\end{align}

This completes the proof of Eq.~\ref{eq:dotM} in the main text.
In the absence of the reservoir, only the unitary part survives, with the matrix $\Lambda$ replaced by the antisymmetric matrix $A$ corresponding to the system Hamiltonian in the Majorana basis.

\section{Spectral analysis for different reservoirs}
\label{app:toymodels}

In this section, we shall derive the couplings entering the Lindblad equation for some simple models and also analyze the resulting spectrum of the Liouvillian from a statistical point of view.

\subsection{The critical Kitaev chain}
We first compute the Lindblad couplings for the critical Kitaev chain. The model is basically the `bath part' of Eq.~\ref{eq:CKC}, with only nearest neighbour couplings among local Majorana operators, i.e.-
\begin{align}
\label{eq:critical_kitaev_chain}
\nonumber
H &= ih\,\sum_{n=1}^{2N} b_n b_{n+1}\\
  &= ih \sum_{m=1}^{N}(b_{2m-1}b_{2m} + b_{2m}b_{2m+1}),
\end{align}
where $\{b_{i},b_{j}\}=\delta_{ij}$. 
We assume that the chain is of even length ($L=2N$) and periodic boundary condition is implemented. To diagonalize it, we first transform to usual fermion operators- 
\begin{align}\nonumber
b_{2m-1} &= \frac{1}{\sqrt{2}}(c_{m}+c^{\dagger}_{m}) \\
b_{2m} &= \frac{i}{\sqrt{2}}(c_{m}-c^{\dagger}_{m}) .
\end{align}
In terms of these operators, the Hamiltonian takes the form-
\begin{equation}
\begin{split}
H=&-h\sum_{n=1}^{N}\left(c^{\dagger}_{m}c_{m}-\frac{1}{2}\right) + \frac{h}{2}\sum_{n=1}^{N}(c^{\dagger}_{m}c_{m+1}+h.c.) \\ &-\frac{h}{2}\sum_{n=1}^{N}(c_{m}c_{m+1} + h.c.) .
\end{split}
\end{equation}
The next step is to go to Fourier space through the transformations- 
\begin{align}\nonumber
c_{m} &= \frac{1}{\sqrt{N}}\sum_{k\in BZ}c_{k}e^{-ikm} \\ 
c^{\dagger}_{m} &= \frac{1}{\sqrt{N}}\sum_{k \in BZ}c^{\dagger}_{k}e^{ikm} .
\end{align}
The 1d Brillouin Zone (BZ) is the interval $[-\pi,\pi)$, where the lattice spacing is set to unity.
The resulting form of the Hamiltonian maybe written compactly using Nambu spinors as- 
\begin{align}\nonumber
H &= \frac{h}{2} \sum_{k \in BZ} \begin{pmatrix} c^{\dagger}_{k} & c_{-k} \end{pmatrix} \mathcal{H}(k)\begin{pmatrix} c_{k} \\ c^{\dagger}_{-k} \end{pmatrix} \\ 
\mathcal{H}(k) &=  \begin{pmatrix} -1+\cos{k} & -i\sin{k} \\ i\sin{k} & 1-\cos{k} \end{pmatrix} .
\end{align}
The spectrum of the model is immediately derivable by solving for the eigenvalues of $h(k)$, which yields two branches- $\zeta_{k} = \pm 2h \abs{\sin(k/2)}$. The matrix $\mathcal{H}(k)$ also has the following discrete symmetries- chiral, time-reversal (TRS) and particle-hole (PHS). These may be summarized by the relations- 
\begin{align}\nonumber
\tau_{x}\mathcal{H}(k)\tau_{x} &= -\mathcal{H}(k) \\ \nonumber
\mathcal{H}^{*}(k) &= \mathcal{H}(-k) \\ 
\tau_{x}\mathcal{H}^{*}(k)\tau_{x} &= -\mathcal{H}(-k) .
\end{align}
Here $\tau_{x}$ is the usual Pauli matrix in the $z$-basis. These symmetries help to relate the eigenvectors for a given branch and positive $k$ to those corresponding to the other branch or negative $k$. For example, the eigenvector for the positive energy branch (for $k>0$) is $(\bar{u}_{k} \quad i\bar{v}_{k})^T$, where- 
\begin{align}\nonumber
\bar{u}_{k} &= \sqrt{\frac{1-\sin(k/2)}{2}} \\ 
\bar{v}_{k} &= \sqrt{\frac{1+\sin(k/2)}{2}} .
\end{align}
Now, using the symmetries mentioned above, we may write the full Hamiltonian (for all $k$) in the diagonal form  as- 
\begin{equation}
H = \sum_{k} 2h\abs{\sin(k/2)} \left( d^{\dagger}_{k}d_{k}-\frac{1}{2}\right) ,     
\end{equation}
with
\begin{align}\nonumber
d_{k} &= u_{k}c_{k}-iv_{k}c^{\dagger}_{-k} , \\ \nonumber
u_{k} &= \sqrt{\frac{1-\abs{\sin(k/2)}}{2}} , \\ 
v_{k} &= sgn(k)\sqrt{\frac{1+\abs{\sin(k/2)}}{2}} .
\end{align}
At this point, we note that $v_{k}$ is discontinuous at $k=0$. So, we have to treat this mode separately (the related $\mathcal{H}(k)=0$). Moreover, we also observe that $J_{ij}$ depends only on this mode, while $\Lambda_{ij}$ excludes it.

For $J_{ij}$, a simple choice of defining the two $k=0$ mode wavefunctions to have constant amplitudes on either the even or odd sites (which are in fact different sublattices), gives us- 
\begin{align}\nonumber
J_{ij} &= \frac{1}{2}\sum_{k} \delta(\zeta_{k}) S_{ijk} \\ \nonumber
&=\frac{w_{i}w_{j}}{2\pi} \int_{-\pi}^{\pi} \frac{dk}{|\zeta^{\prime}_{k}|} \delta(k) \\ 
&= \begin{cases} 
\dfrac{w_{i}w_{j}}{2\pi h}, &\textrm{if}\;i,j \;\textrm{are odd,}\\ 
\\
\dfrac{w_{i}w_{j}}{2\pi h}, &\textrm{if}\;i,j \;\textrm{are even,}\\ 
\\
0, &\textrm{otherwise} .
\end{cases}
\end{align}

In order to derive $\Lambda_{ij}$ couplings, we need to relate the diagonal modes to the local Majorana modes we started with.
First, we introduce the canonical Majorana modes through the relations- 
\begin{align}\nonumber
d_{k} &= \frac{1}{\sqrt{2}}(\tilde{b}_{2p-1} - i\tilde{b}_{2p}) \\ 
d^{\dagger}_{k} &= \frac{1}{\sqrt{2}}(\tilde{b}_{2p-1}+i\tilde{b}_{2p}) .
\end{align}
In terms of these modes, the Hamiltonian looks like- 
\begin{equation}
H = i\sum_{p=1}^{N}\zeta_{p}(k)\tilde{b}_{2p}\tilde{b}_{2p-1} .
\end{equation}
After some manipulations, the `inverse transformations' relating the local Majorana modes to these canonical ones maybe written as- 
\begin{equation*}
\begin{split}
b_{2n-1} = \frac{1}{\sqrt{N}}\sum_{p=1}^{N}\;&[u_{p}(k)\cos(kn)+v_{p}(k)\sin(kn)]\,\tilde{b}_{2p-1}\\ +&[v_{p}(k)\cos(kn)-u_{p}(k)\sin(kn)]\,\tilde{b}_{2p} ,  
\end{split}
\end{equation*}
where we have changed notation slightly and redefined $u_{k}$ and $v_{k}$ as $u_{p}(k)$ and $v_{p}(k)$.

One can now read off the matrix elements of $\bar{O}^{T}$ as
\begin{align}\nonumber
\bar{O}^{T}_{2n-1,2p-1} &= \frac{1}{\sqrt{N}}(u_{p}(k)\cos(kn)+v_{p}(k)\sin(kn)) \\ \nonumber 
\bar{O}^{T}_{2n-1,2p} &= \frac{1}{\sqrt{N}}(v_{p}(k)\cos(kn)-u_{p}(k)\sin(kn)) \\ \nonumber 
\bar{O}^{T}_{2n,2p-1} &= \frac{1}{\sqrt{N}}(v_{p}(k)\cos(kn)+u_{p}(k)\sin(kn)) \\ 
\bar{O}^{T}_{2n,2p} &= \frac{1}{\sqrt{N}}(u_{p}(k)\cos(kn)-v_{p}(k)\sin(kn)) .
\end{align}
The `bare' tunneling matrix can be taken as purely local. Then, if $i$ and $j$ are both even or odd ($i=2n-1, j=2m-1$ or $i=2n, j=2m$), one finds-
\begin{equation}
A_{ijp} = -\frac{w_{i}w_{j}}{N}\sin[k(n-m)] , 
\end{equation}
where $w_{i}$ and $w_{j}$ are real tunneling amplitudes on sites $i$ and $j$ respectively.
Alternatively if one is even and the other is odd ($i=2n-1, j=2m$ or $i=2n, j=2m-1$), we obtain-
\begin{equation}
A_{ijp} = -\frac{w_{i}w_{j}}{N}\sqrt{1-\abs{\sin(k/2)}}\,\sgn(k)\sin[k(n-m)] . 
\end{equation}

Substituting this expression directly into the formula for $\Lambda_{ij}$ yields- 
\begin{align}\nonumber
\Lambda_{ij} &=  \frac{1}{2} \sum_{p} \zeta_{p}(k)^{-1} A_{ijp} \\
&= \begin{cases} 
-\dfrac{w_{i}w_{j}}{2h}, &\textrm{if}\;i\;\textrm{is odd},j \;\textrm{is even,}\\
\\
-\dfrac{w_{i}w_{j}}{2h}, &\textrm{if}\;i\;\textrm{is even},j \;\textrm{is odd,}\\ 
\\
0, &\textrm{otherwise} .
\end{cases}
\end{align}

\subsection{Continuum fermion models in various dimensions}
Next, we consider, as the reservoir, a free electron gas in $d$ dimensions with $d=1,2,3$. The corresponding Hamiltonian reads- 
\begin{equation}
H = \sum_{{\mathbf k}} \zeta_{\mathbf k} c^{\dagger}_{\mathbf k}c_{\mathbf k}, \quad \zeta_{\mathbf k} = \frac{k^2}{2m}-\mu .
\end{equation}
We define the canonical Majorana operators as- 
\begin{align}\nonumber
c_{\mathbf k} &= \frac{1}{\sqrt{2}}(b_{\mathbf{k},1} - ib_{\mathbf{k},2}) \\
c^{\dagger}_{\mathbf k} &= \frac{1}{\sqrt{2}}(b_{\mathbf{k},1} + ib_{\mathbf{k},2}) .
\end{align}
The local Majorana operators may be defined similarly. 
However, since 
\begin{equation}
c({\mathbf r}) = \frac{1}{(2\pi)^{d/2}}\int d{\mathbf k} e^{-i{\mathbf k}.{\bf r}}c_{\mathbf k} ,     
\end{equation}
we have the following relations- 
\begin{align}\nonumber
b_{1}({\mathbf r}) &= \frac{1}{(2\pi)^{d/2}} \int d{\mathbf k}\,[\cos({\mathbf k}\cdot{\mathbf r})b_{\mathbf{k},1} - \sin({\mathbf k}\cdot{\mathbf r})b_{\mathbf{k},2}]\,, \\  
b_{2}({\mathbf r}) &= \frac{1}{(2\pi)^{d/2}} \int d{\mathbf k}\,[\sin({\mathbf k}\cdot{\mathbf r})b_{\mathbf{k},1} + \cos({\mathbf k}\cdot{\mathbf r})b_{\mathbf{k},2}]\,.
\end{align}
Now, one may read off the matrix elements of $\bar{O}^{T}$. If a computational Majorana couples to $b_{1}({\bf r_{j}})$, we have- 
\begin{align}\nonumber
\bar{O}^{T}_{j,2{\mathbf p}-1} &= \frac{1}{(2\pi)^{d/2}}\cos(\mathbf{k}\cdot\mathbf{r}_j) \\
\bar{O}^{T}_{j,2{\mathbf p}} &= -\frac{1}{(2\pi)^{d/2}}\sin(\mathbf{k}\cdot\mathbf{r}_j) . 
\end{align}
Otherwise, if it couples to $b_{2}({\bf r_{j}})$, then- 
\begin{align}\nonumber
\bar{O}^{T}_{j,2{\mathbf p}-1} &= \frac{1}{(2\pi)^{d/2}}\sin(\mathbf{k}\cdot\mathbf{r}_j) \\  
\bar{O}^{T}_{j,2{\mathbf p}} &= \frac{1}{(2\pi)^{d/2}}\cos(\mathbf{k}\cdot\mathbf{r}_j) .
\end{align}

The `bare' tunneling matrix in this case couples one computational Majorana mode to a local fermion. Hence, one has couplings to both $b_{1}({\mathbf r}_{j})$ and $b_{2}({\mathbf r}_{j})$, with strengths proportional to $\im(w_{j})$ and $\re(w_{j})$ respectively. After simplification, one finds the Lindblad couplings as:
\begin{align}\nonumber
\Lambda_{ij} &=\im(w_i w^*_j)\,\textrm{p.v.}\,\int \frac{d\mathbf{k}}{(2\pi)^d} \frac{\cos[\mathbf{k}\cdot(\mathbf{r}_i-\mathbf{r}_j)]}{\zeta_{\mathbf{k}}} , \\ \nonumber
J_{ij} &=\re(w_i w^*_j)\, \int \frac{d\mathbf{k}}{(2\pi)^d} \,\delta(\zeta_{\mathbf{k}})\,\cos[\mathbf{k}\cdot(\mathbf{r}_i-\mathbf{r}_j)] , \\ 
\zeta_{\mathbf{k}} &= k^2 /2m - \mu .
\end{align}

The integrals can be expressed in terms of the real and imaginary parts of the fermion Green's functions at zero energy, and can be found e.g. in Ref.~\cite{economou}. In terms of the quantities defined above, leading us to Eqs.~\eqref{eq:Js_and_Lambdas}.

\section{Comments on the reservoir correlation function}
\label{app:reservoir_correlation_functions}

In this section, to understand the content of the Born-Markov approximation, we calculate the reservoir correlation function $C_{ij}(t)$ explicitly for the one-dimensional model of continuum fermions. The expressions for $S_{ijk}$ and $A_{ijk}$ are straightforward to derive-
\begin{align}\nonumber
S_{ijk} &= 2\re(w_i w^*_j)\, \frac{1}{(2\pi)}\cos[k(x_i-{x}_j)] \\ 
A_{ijk} &= 2\im(w_i w^*_j)\, \frac{1}{(2\pi)}\cos[k(x_i-{x}_j)] .
\end{align}
Here $x_i$ and $x_j$ denote the positions of the $i$-th and $j$-th Majorana modes respectively.
After putting in the above expressions in Eq.~\ref{eq:reservoir correlation} and simplifying for zero temperature ($\beta\rightarrow\infty$), we obtain
\begin{equation}
C_{ij}(t) = \frac{w_{i}w^*_{j}}{2\pi}\sum_{k}e^{ik(x_i-x_j)}e^{-i(k^2/2m-\mu)t} .
\end{equation}
Converting the sum to an integral, we may do the same exactly as it's a simple Gaussian one- 
\begin{align}\nonumber
C_{ij}(t) &= w_{i}w^*_{j} \int \frac{dk}{2\pi} e^{ik(x_i-x_j)}e^{-i(k^2/2m) t} \\ &= w_{i}w^*_{j}\frac{e^{im(x_i-x_j)^2/(2t)}e^{i\mu t}}{\sqrt{2\pi}} \frac{1}{\sqrt{it/m}} .
\end{align}
This is an oscillatory function with a power law ($\propto1/\sqrt{t}$) decay. However, the equation of motion for $\rho$ contains only integrals of this correlation function over time. These indeed have a dominant contribution for small times, after which they merely oscillate about a mean value.

This lends validity to the Born-Markov approximation we invoked to derive the effective Lindblad equation~\eqref{eq:lindblad}. We also comment that in presence of potential disorder, the correlation function shall also contain an exponential decay, with a time scale proportional to $1/W$, where $W$ is the disorder strength. Provided this scale is much shorter than the timescales corresponding to the system-reservoir coupling, again Markov approximation may be justified.

\bibliography{references}

\end{document}